\documentclass[twocolumn,amssymb, nobibnotes, aps, prd, superscriptaddress]{revtex4-1}

\usepackage{graphicx}
\usepackage{amsmath, amsthm, amssymb}
\usepackage{mathdots}
\usepackage{mathtools}
\usepackage{float}
\usepackage{tikz-cd} 
\usepackage{tikz}
\usetikzlibrary{
  arrows.meta,                        
  backgrounds,                        
  ext.paths.ortho,                    
  ext.positioning-plus,               
  ext.node-families.shapes.geometric, 
  calc}                               
\usepackage{pgfplots} 
\usepgfplotslibrary{patchplots}
\usepgfplotslibrary{groupplots}
\pgfplotsset{compat=1.5.1}
\usetikzlibrary{positioning}
\usetikzlibrary{shapes}
\usetikzlibrary{shapes.geometric}
\usetikzlibrary{decorations.text}
\usetikzlibrary{decorations.pathreplacing,shapes.misc}
\usetikzlibrary{patterns}
\usetikzlibrary {patterns.meta} 
\usetikzlibrary{matrix}
\usepackage{caption,subcaption}
\usepackage{nicematrix}
\theoremstyle{definition}
\newtheorem{definition}{Definition}[section]
\newtheorem{theorem}{Theorem}[section]

\newtheorem{lemma}{Lemma}[theorem]

\usepackage{apptools} 
\AtAppendix{\counterwithin{lemma}{section}}
\AtAppendix{\counterwithin{definition}{section}}
\AtAppendix{\counterwithin{theorem}{section}}
\AtAppendix{\counterwithin{conjecture}{section}}
\usepackage[ruled,lined]{algorithm2e}
\usepackage{algorithmic}
\renewcommand{\algorithmiccomment}[1]{\bgroup\hfill//~#1\egroup}

\newcommand\rank{rank\:}
\newcommand\conn{Conn}
\newcommand\Fid{Id}
\newcommand\quotient[2]{
	\mathchoice
	{
		\text{\raise1ex\hbox{$#1$}\Big/\lower1ex\hbox{$#2$}}%
	}
	{
		#1\,/\,#2
	}
	{
		#1\,/\,#2
	}
	{
		#1\,/\,#2
	}
}

\begin{document}

\title{Rigidity condition for gluing two bar-joint rigid graphs embedded in $\mathbb{R}^d$}
\author{Kyungeun Kim}
\email{kkim10@syr.edu}
\affiliation{Physics Department, Syracuse University, Syracuse, NY 13244 USA}
\author{J. M. Schwarz}
\email{jmschw02@syr.edu}
\affiliation{Physics Department, Syracuse University, Syracuse, NY 13244 USA}
\affiliation{Indian Creek Farm, Ithaca, NY 14850 USA}
	
\date{\today}
\begin{abstract}
How does one determine if a collection of bars joined by freely rotating hinges cannot be deformed without changing the length of any of the bars? In other words, how does one determine if a bar-joint graph is rigid? This question has been definitively answered using combinatorial rigidity theory in two dimensions via the Geiringer-Laman Theorem. However, it has not yet been answered using combinatorial rigidity theory in higher dimensions, given known counterexamples to the trivial dimensional extension of the Geiringer-Laman Theorem. To work towards a combinatorial approach in dimensions beyond two, we present a theorem for gluing two rigid bar-joint graphs together that remain rigid. When there are no overlapping vertices between the two graphs, the theorem reduces to Tay's theorem used to identify rigidity in body-bar graphs. When there are overlapping vertices, we rely on the notion of pinned rigid graphs to identify and constrain rigid motions. This theorem provides a basis for an algorithm for recursively constructing rigid clusters that can be readily adapted for computational purposes. By leveraging Henneberg-type operations to grow a rigid (or minimally rigid) graph and treating simplices-where every vertex connects to every other vertex-as fundamental units, our approach offers a scalable solution with computational complexity comparable to traditional methods. Thus, we provide a combinatorial blueprint for algorithms in multi-dimensional rigidity theory as applied to bar-joint graphs.
\end{abstract}
		
\maketitle
		
\section{Introduction}
The concept of rigidity and the conditions for determining it have deep historical roots, tracing back to the works of Euler (1707–1783) and Lagrange (1736–1813). Euler, in one of his books published posthumously, wrote that ``Closed surfaces, such as spheres, are rigid and do not change shape, while open surfaces, like hemispheres, can be altered,” highlighting the rigidity of closed surfaces \cite{Euler1862}. Lagrange addressed the equilibrium of bar linkages in his 1811 book, {\it M\'ecanique analytique}, in the chapter titled ``On the equilibrium of several forces applied to a system of bodies considered as points and connected by strings or rods” \cite{Lagrange1811}. The chapter explores how forces and constraints influence the rigidity of mechanical systems.

A key development in combinatorial rigidity theory came in 1864 ``Maxwell counting” in three dimensions, where each point in a system with $s$ points and $e$ connections has three equilibrium equations, resulting in $3s$ equations for the entire system. Additionally, six equilibrium conditions must be satisfied due to the balance of action and reaction. Therefore, if $e = 3s - 6$, any external force will produce stresses in the system's components to become rigid. Both Beiringer and Laman formalized this concept for generic, two-dimensional (2D) bar-joint graphs in 1927 and in 1970 respectively, to provide a rigorous condition for rigidity~\cite{Laman1970,Geiringer1927}. Here, generic means that there are no algebraic relations between the positions of the vertices. Specifically, it asserts that a graph is rigid in two dimensions if (a) it meets edge-counting criteria $e = 2s - 3$ and (b) the same holds for every subgraph with at least two vertices. Jacobs and Hendrickson~\cite{JACOBS1997} later translated these conditions into an algorithm known as the pebble game, which is efficient for 2D bar-joint graphs and since extended to frictional rigidity~\cite{Henkes2016,Liu2019,Liu2021,Van2024}.

Determining the rigidity of three-dimensional (3D) bar-joint graphs based solely on connectivity remains an unresolved problem in rigidity theory. While Laman’s theorem provides a solution in 2D, its extension to 3D does not fully address all cases, as noted by Thorpe and Duxbury \cite{Thorpe2002} and Liberti {\it et al.} \cite{Liberti2016}. Cheng, {\it et al.} \cite{CHENG2014} offered a partial solution for 3D combinatorial bar-joint graph rigidity using graph covering techniques. Several other approaches exist, including the work of Sitharam {\it et al.} \cite{Sitharam2004}, applying the concept of "module-rigidity" to check the rigidity of 3D graphs, and another relying on a combination of the pebble-game algorithm, rigidity matrices, and hinge-checking techniques \cite{Chubynsky2007}. 

\begin{figure}
		\captionsetup{singlelinecheck = false, justification=raggedright}
		\begin{center}
\begin{tabular}{c}
     \includegraphics[width=0.4\textwidth]{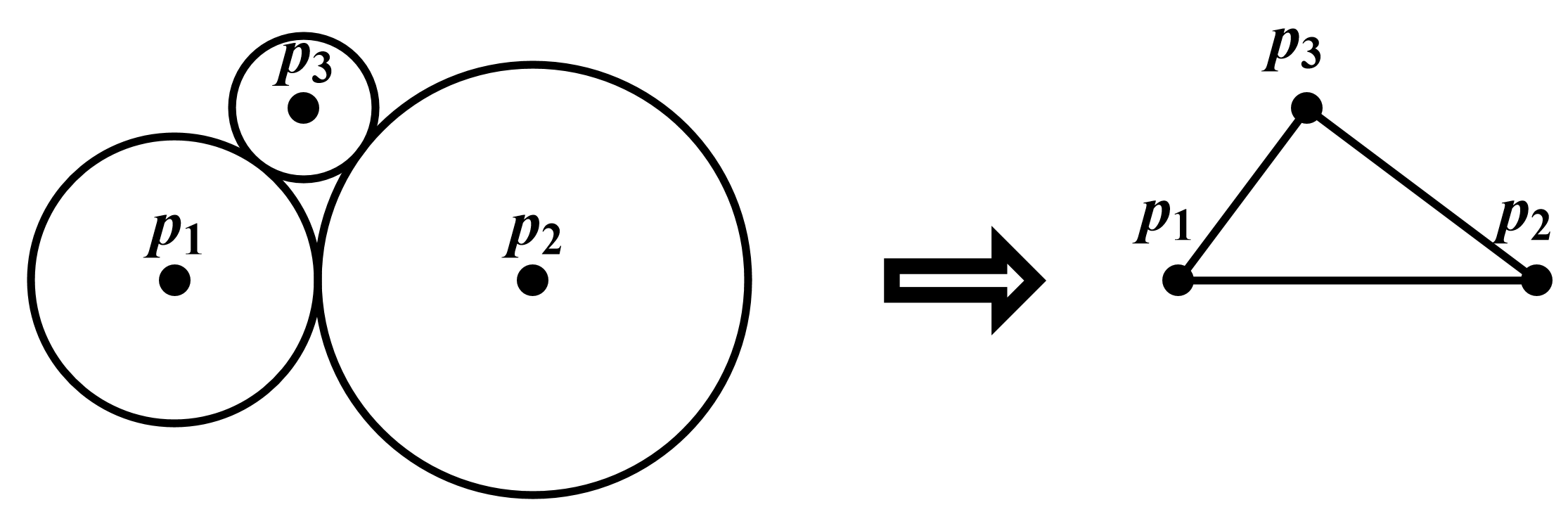}\\
    (a) Circle packing converted to a bar-joint graph\\
     \includegraphics[width=0.45\textwidth]{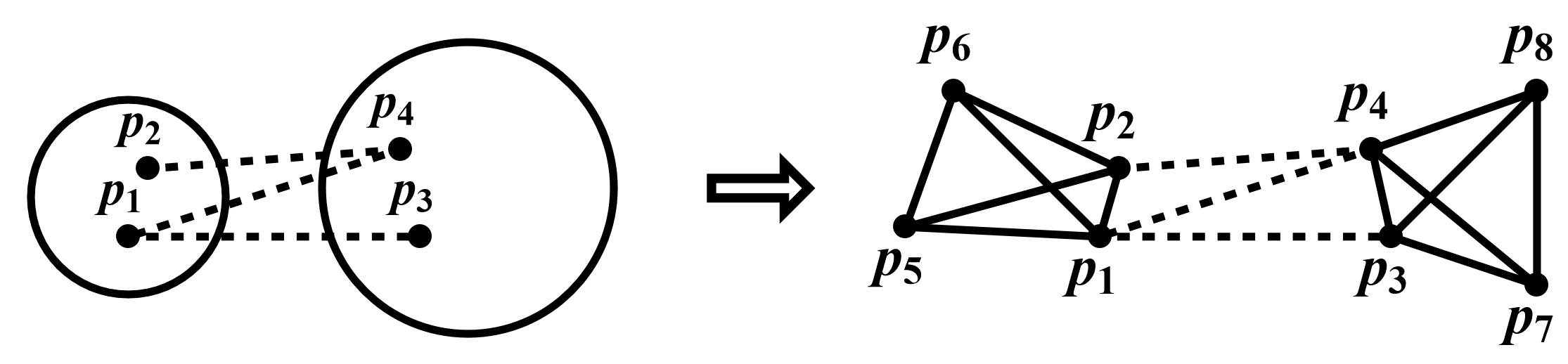}\\
(b) Body-bar graph converted to a bar-joint graph\\
\end{tabular}
	\caption{{\it Examples of converting circle packings and body-bar graphs to bar-joint graphs.} (a) A fully packed circle configuration can be represented as a bar-joint graph, where edges correspond to the distances between circle centers. (b) A body can be represented as a globally rigid graph in the bar-joint framework.}
  \label{fig-body-bar}
		\end{center}
	\end{figure}

Other combinatorial rigidity theory advances focus on body-bar rigidity, where multiple rigid bodies are connected by bars \cite{White1987,schulze2014}. Tay established the formal condition for body-bar rigidity through Tay's theorem, proving that a multi-graph can be realized as a rigid linkage of rigid bodies in $d$-dimensional space if and only if it contains $\frac{d(d+1)}{2}$ edge-disjoint spanning trees \cite{TAY1984}. Body-bar graphs, as shown in Fig. \ref{fig-body-bar}(b), abstractly represent bodies and their connections, functioning similarly to linkages between multiple bodies. Body-bar rigidity has applications in molecular and protein studies \cite{Sljoka2022}, with Tay’s theorem \cite{TAY1984} providing a solid framework for analyzing these systems. Additionally, the study of body-hinge networks explores the rigidity of graphs connected by hinges, providing further insights beyond traditional bar connections \cite{JORDAN2016}. 

Another significant extension of rigidity theory is the concept of "pinned rigid graphs," which describe the movement constraints in pinned networks or mechanical linkages \cite{Jordan2010,Shai2013}. The rigidity of a graph directly relates to whether its solution space is finite or infinite. A finite solution space implies a limited number of distinct realizations for the system, meaning the structure is rigid and only a few configurations are possible \cite{Borcea04,Grasegger2018,JACKSON2019,Koutschan2021}. Conversely, an infinite solution space does not imply unrestricted freedom, but rather that the system exhibits constrained movement, where each vertex or body is restricted to a unique trajectory or a limited range of positions. This leads to an infinite number of realizations but within specific bounds. Understanding whether a system's solution space is finite or infinite is crucial, as it determines whether the system is fully rigid or capable of limited movement.

To more closely approach a combinatorial algorithm for determining the rigidity of generic bar-joint graphs in three dimensions and beyond, we propose a novel algorithm, integrating concepts from Tay's theorem and pinned rigid graphs, to combinatorially determine how to connect to rigid clusters such that the connected structure remains rigid. It is a bottom-up approach grounded in the simplex substructures of the graph.  We anticipate that our algorithm, with its accompanying proof, will have implications across a broad range of systems. For instance, Asimow and Roth \cite{Asimow1978,ASIMOW1979} have extended rigidity theory to more general graphs with edge functionality so that rigidity theory extends to systems such as frictionless and frictional circle packings/spheres~\cite{Schramm1991,CONNELLY2008,bonsante2023,connelly2024,Henkes2016,Liu2019,Liu2021} and robotic assemblies \cite{Briot2019,BRUUN2022}, communication networks between robots \cite{Krick2009,Agarwal2013,Zelazo2015,Riyas2020}, and to other systems, such as glasses~\cite{Thorpe2002}, granular packings~\cite{Wyart2005,Goodrich2013}, and molecules \cite{Whiteley2005}.

Our manuscript is organized as follows: In Sec. \ref{sec02}, we introduce key concepts, including the basic notation of rigid graphs. Next, in Sec. \ref{sec03}, we discuss the rigid construction of graphs, illustrating the connection between pinned rigid graphs and rigid structures in two and three dimensions, and generalizing these results to higher dimensions. In Sec. \ref{sec04}, we propose an algorithm that minimizes reliance on the rigidity matrix for more efficient computation. We conclude with a discussion of the potential impact our work in terms of how our approach provides a scalable solution with computational complexity comparable to traditional methods.

\section{Basic Concepts}\label{sec02}
In this section, we introduce key notations and concepts related to graph rigidity. Although a graph can be described abstractly by naming its vertices and edges, it lacks specific spatial coordinates until it is embedded in a certain space. To formalize this, we define the framework of a graph as follows:

\begin{definition} A framework $G(V,E)$ of a graph $G$ is defined by a set of vertices $V=\{v_1,\dots,v_n\}$ (denoted $V(G)$) and corresponding edges $E=\{e_1,\dots,e_n\}$ (denoted $E(G)$) that connect these vertices (for example, $e_1=(v_1,v_2)$). The notation $|V|$ represents the number of vertices, and $|E|$ represents the number of edges. \end{definition}

Next, we define a complete graph, which contains all possible connections between its vertices.

\begin{definition}[Section 2.1 \cite{ComRig}] Let $V$ be a fixed set of $n$ elements, and let $K$ be the collection of all unordered pairs of elements in $V$. Then $(V, K)$ is called a complete graph. \end{definition}

We denote a complete graph with $n$ vertices as $K_n$, which has $\binom{n}{2} = \frac{n(n-1)}{2}$ edges. The figure below (Fig. 2) illustrates complete graphs for different values of $n$.
\begin{figure}[H]
    \centering
\begin{tikzpicture}[scale=0.4, transform shape]
		\node [fill=black,circle] (1) at (-1, 3.5) {};
		\node [fill=black,circle] (2) at (-2, 1) {};
		\node [fill=black,circle] (3) at (2.5, 3.75) {};
		\node [fill=black,circle] (4) at (1, 1.25) {};
		\node [fill=black,circle] (5) at (3.5, 1.25) {};
		\node [fill=black,circle] (6) at (9, 3.75) {};
		\node [fill=black,circle] (7) at (5.5, 1.5) {};
		\node [fill=black,circle] (8) at (8.75, 0.75) {};
		\node [fill=black,circle] (9) at (5.75, 4.25) {};
		\node [] (12) at (-1.5, -0.25) {};
		\node [] (13) at (-1.5, -0.25) {\LARGE$K_2$};
		\node [] (14) at (2.25, -0.25) {\LARGE$K_3$};
		\node [] (15) at (7, -0.25) {\LARGE$K_4$};
		\node [fill=black,circle] (16) at (15, 3.5) {};
		\node [fill=black,circle] (17) at (11.25, 1) {};
		\node [fill=black,circle] (18) at (14.25, 0.75) {};
		\node [fill=black,circle] (19) at (10.75, 3.25) {};
		\node [] (20) at (12.5, -0.25) {\LARGE$K_5$};
		\node [fill=black,circle] (21) at (12.75, 4.75) {};
		\draw [-,thick](1) to (2);
		\draw [-,thick](3) to (4);
		\draw [-,thick](4) to (5);
		\draw [-,thick](5) to (3);
		\draw [-,thick](9) to (7);
		\draw [-,thick](9) to (8);
		\draw [-,thick](9) to (6);
		\draw [-,thick](6) to (8);
		\draw [-,thick](6) to (7);
		\draw [-,thick](7) to (8);
		\draw [-,thick](19) to (17);
		\draw [-,thick](19) to (18);
		\draw [-,thick](19) to (16);
		\draw [-,thick](16) to (18);
		\draw [-,thick](16) to (17);
		\draw [-,thick](17) to (18);
		\draw [-,thick](21) to (19);
		\draw [-,thick](21) to (16);
		\draw [-,thick](21) to (17);
		\draw [-,thick](21) to (18);
\end{tikzpicture}
    \caption{{\it A complete graph for different $n$s.}}
\end{figure}
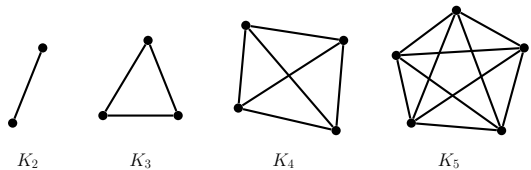

\begin{figure}
		\captionsetup{singlelinecheck = false, justification=raggedright}
		\begin{center}
     \includegraphics[width=0.45\textwidth]{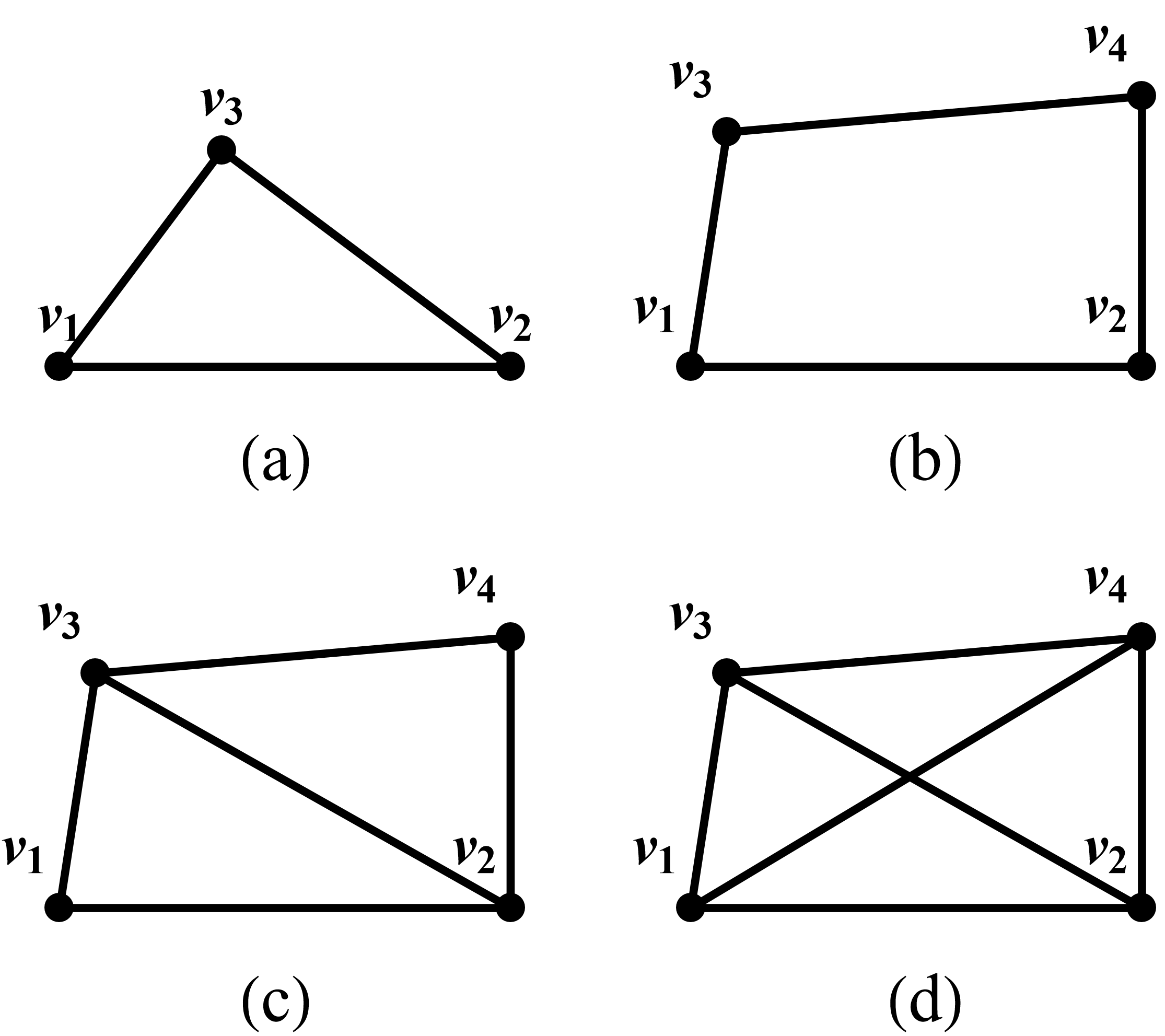}\\
    \caption{{\it Examples of rigid and flexible bar-joint graphs with pinned points.} (a) A triangle (minimally rigid). (b) A tetragon (flexible). (c) A tetragon with one additional edge (or two neighboring triangles) (minimally rigid). (d) A tetragon with two additional edges (globally rigid).}
  \label{fig-rigid_ex}
		\end{center}
	\end{figure}

In this abstract structure, we do not assume specific edge lengths, though they must be nonzero. However, if we assign arbitrary lengths $l_i$ to each edge, we can illustrate the concept of rigidity. For example, consider Fig. \ref{fig-rigid_ex} (a), where, for convenience, we arbitrarily fix vertex $v_1$ at $(0,0)$, and place vertex $v_2$ at $(l_2, 0)$. The position of vertex $v_3$ can then be determined using two circle equations based on the edge lengths $l_1=(v_1,v_3)$ and $l_3=(v_2,v_3)$, resulting in two possible positions for $v_3$ (marked as $\bigtriangleup$ and $\bigtriangledown$). This configuration forms a rigid triangle.

In contrast, the tetragon shown in Fig. \ref{fig-rigid_ex} (b) has four variables (the coordinates of $v_3$ and $v_4$), but only three independent equations, resulting in one degree of freedom, allowing the structure to flex. Thus, the tetragon is flexible. By adding an additional edge, as shown in Fig. \ref{fig-rigid_ex} (c), we form two connected triangles. This creates more constraints, leading to $2^n$ possible configurations, where $n$ is the number of non-overlapping triangles \cite{Borcea04}. Finally, adding one more edge results in only one possible shape (up to reflection), making the structure globally rigid. Both (a) and (c) are minimally rigid, meaning that removing any edge would render the structure flexible.

 Let us now assume that we do not know the specific coordinates of each vertex $v_i$ in $G$, but we know that the dimension of each $x_i$ corresponds to the vertices, and the $l_i$ represent symbolic lengths of the edges. The framework remains abstract, the vertices $x_i$ are unknown, and the lengths $l_i$ serve as parameters representing the constraints between the vertices. We can model the structure as a matroid and express the relationships between vertices and edges using a matrix, which captures the tension in the edges based on their lengths \cite{CALLADINE1978}. These linear relations can be represented as a matrix, called the rigidity matroid $\mathcal{R}(G)=\mathcal{R}(G,x)$, where $x_i$ refers to the symbolic positions of the vertices, but their actual coordinates remain unknown.

For example, for the triangle in Fig. \ref{fig-rigid_ex} (a), we have the following relations for the edges and vertices:

\setlength{\arraycolsep}{1pt}
\[
\begin{bNiceMatrix}[first-row, first-col]
\CodeBefore
\rectanglecolor{blue!15}{1-1}{1-1}
\rectanglecolor{blue!15}{1-3}{1-3}
\rectanglecolor{blue!15}{2-1}{2-2}
\rectanglecolor{blue!15}{3-2}{3-3}
\Body  
&v_1 &v_2&v_3\\
e_1=(v_1,v_3)&x_1-x_3&0&x_3-x_1\\
e_2=(v_1,v_2)&x_1-x_2&x_2-x_1&0\\
e_3=(v_2,v_3)&0&x_2-x_3&x_3-x_2\\
\end{bNiceMatrix} 
\].
Here, the $x_i$s are not specific coordinates but instead symbolic representations of the vertices, while the edges encode the relationship between these vertices.

The matroid does not require knowledge of actual vertex positions in space—only the connections (or linkages) between the vertices. If the coordinates were known, the matroid could then be converted to a rigidity matrix $\mathcal{R}(G,p)$, where $p_i$ represents actual positions of the vertices.
\[
\begin{bNiceMatrix}[first-row, first-col]
\CodeBefore
\rectanglecolor{blue!15}{1-1}{1-1}
\rectanglecolor{blue!15}{1-3}{1-3}
\rectanglecolor{blue!15}{2-1}{2-2}
\rectanglecolor{blue!15}{3-2}{3-3}
\Body  
&v_1 &v_2&v_3\\
e_1=(v_1,v_3)&p_1-p_3&0&p_3-p_1\\
e_2=(v_1,v_2)&p_1-p_2&p_2-p_1&0\\
e_3=(v_2,v_3)&0&p_2-p_3&p_3-p_2\\
\end{bNiceMatrix} 
\].
The rigidity matrix describes the linear dependence of the vertex positions on the lengths of the edges. This matrix allows us to determine whether a framework is infinitesimally rigid (locally rigid but flexible at larger scales) or globally rigid (rigid at all scales) by comparing the rank of the matrix with the number of edges. Some papers, such as \cite{Jordan2022,lew2023}, use normalized coordinates $\frac{p_i - p_j}{||p_i - p_j||}$ instead of $p_i - p_j$. In this work, we adopt the notations from \cite{Thorpe2002,Whiteley2005}. The rank of the rigidity matrix has the following property:

\begin{lemma}[\protect{Lemma 1.1}{\cite{Jackson05}}]\label{Srankmatrix}
    Let $(G,p)$ be a framework where $G$ is positioned in $\mathbb{R}^d$ with vertices having $d$-dimensional coordinate vectors $p_i$. Then the rank of $\mathcal{R}(G,p)$ is at most $S(n,d)$, where $n = |V(G)|$ and
    \begin{align}
        S(n,d) = \begin{cases}
            nd - \binom{d+1}{2}, & \text{if } n \geq d+2, \\
            \binom{n}{2}, & \text{if } n \leq d+1.
        \end{cases}
    \end{align}
\end{lemma}

Thus, $G(V,E)$ is rigid in $\mathbb{R}^d$ if $\rank \mathcal{R}(G,p) = S(n,d)$. Notice that if $d \leq n \leq d+1$, then $nd - \binom{d+1}{2} = \binom{n}{2}$, and if $n < d$, then $G$ is not rigid unless $G$ is a complete graph. This also implies that all edges are independent (i.e., independent row vectors in the rigidity matrix) to satisfy $|E(G)| = S(n,d)$. Thus, we define:

\begin{definition}[\protect{Section 11.1}{\cite{Whiteley95}}]\label{dind}
    The framework $(G,p)$ is $d$-independent if its edge set is independent in $\mathcal{R}(G,p)$.
\end{definition}

The graph is said to be infinitesimally rigid if the rank of the rigidity matrix is full:

\begin{theorem}[\protect{Theorem 2.1}{\cite{Sch10}}]
    A framework $(G,p)$ in $\mathbb{R}^d$ is infinitesimally rigid if and only if either $\rank \mathcal{R}(G,p) = S(n,d)$ or $G$ is a complete graph $K_n$.
\end{theorem}

Since a graph $G$ is rigid if it is infinitesimally rigid, $G(V,E)$ is rigid in $\mathbb{R}^d$ if $\rank \mathcal{R}(G,p) = S(n,d)$. Although there are slightly different notations defining rigidity of a graph, from here on, we will consider a graph to be rigid if it has the full rank of the rigidity matrix. Thus, the graph is rigid if $\rank \mathcal{R}(G,p) = S(n,d)$.

Next, we will consider the graph with pinned vertices.

\begin{definition}[\protect{Section 3.1}{\cite{Shai2013}}]\label{rrankprigid}
    For a graph $G(V,E)$ in $\mathbb{R}^d$, we can define a pinned rigid graph $G(I,P,E) = (\widetilde{G}, p)$, where $I$ is the set of inner vertices, $P$ is the set of pinned vertices, and $E$ is the set of edges, with each edge $e \in E$ having at least one endpoint in $I$.
\end{definition}

\begin{figure}
		\captionsetup{singlelinecheck = false, justification=raggedright}
		\begin{center}
\begin{tabular}{c}
     \includegraphics[width=0.45\textwidth]{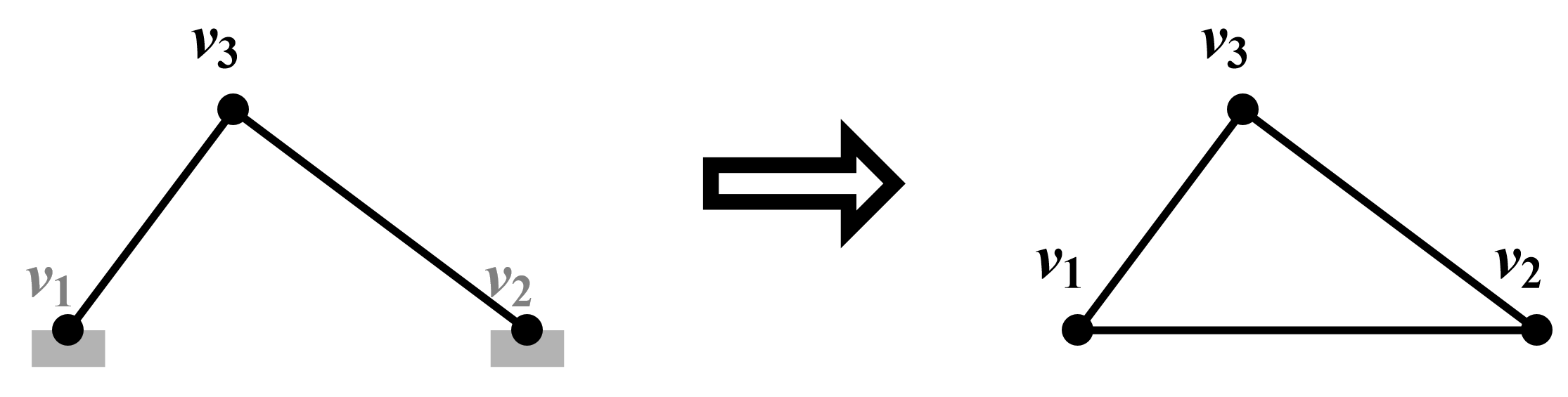}\\
    (a) Two-bar linkage converted to a bar-joint graph\\
     \includegraphics[width=0.45\textwidth]{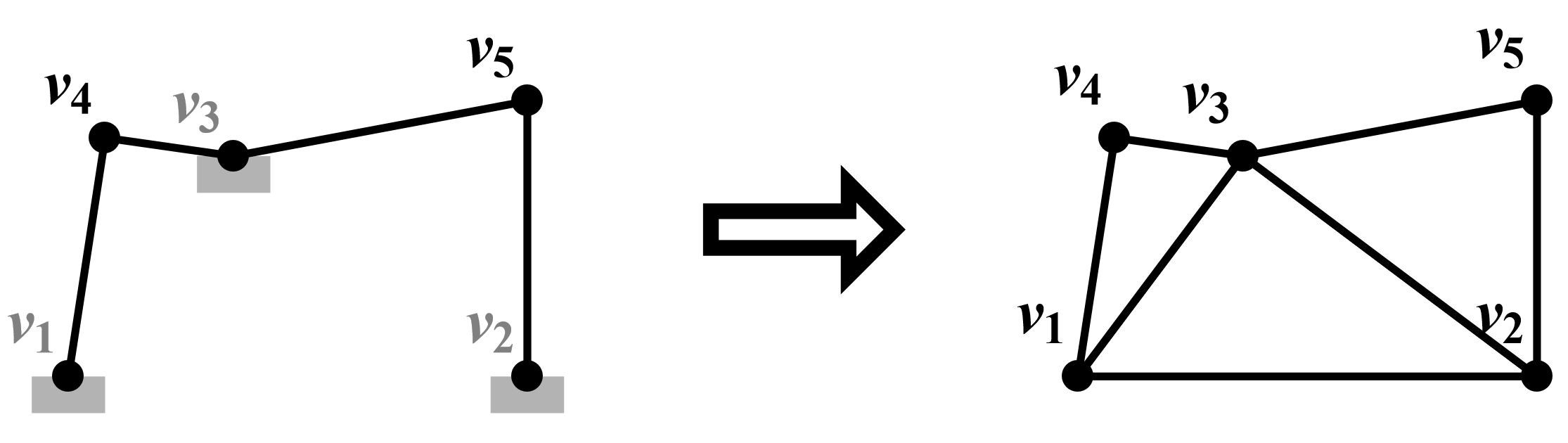}\\
    (b) Pinned rigid graph converted to a bar-joint graph\\
\end{tabular}

			\caption{{Examples of converting graphs with pinned points (gray shaded vertices) to bar-joint graphs.} (a) A two-bar linkage, fixed at points $v_1$ and $v_2$, can be converted to a bar-joint graph with a bar connecting $v_1$ and $v_2$. (b) A pinned rigid graph with three pinned points $v_1,v_2,v_3$ converted to a bar-joint graph.}
  \label{fig-pinned_ex}
		\end{center}
	\end{figure}

Note that a pinned graph can be thought of as a bar linkage fixed in space. Fig. \ref{fig-pinned_ex} shows examples of pinned graphs. For instance, in (a) on the left, $v_1$ and $v_2$ are pinned points, and $v_3$ is an inner vertex. Thus, we can write $P = \{v_1, v_2\}$ and $I = \{v_3\}$. If the number of pinned points is equal to or greater than the number of dimensions, then we can convert the pinned points into a complete (or globally rigid) graph, as shown in the right figure in \ref{fig-pinned_ex} (a),(b). 

Similar to a rigid graph, there exists a pinned rigidity matrix in $\mathbb{R}^d$ for a pinned rigid graph.

\begin{definition}[\protect{Section 3.1}{\cite{Shai2013}}]\label{defdrigid}
    A framework $(\widetilde{G}, p)$ is pinned $d$-rigid if the only infinitesimal motion is the zero motion; equivalently, if the pinned rigidity matrix $\mathcal{R}(\widetilde{G}, p)$ has full rank $d|I|$.
\end{definition}

The pinned rigidity matrix can be defined similarly to the rigidity matrix. For example, considering Fig. \ref{fig-pinned_ex} (b), the pinned rigidity matrix $\mathcal{R}(\widetilde{G}, p)$ can be written as:

\[
\begin{bNiceMatrix}[first-row, first-col]
& v_4 & v_5 \\
e_1 = (v_1, v_4) & p_4 - p_1 & 0 \\
e_2 = (v_3, v_4) & p_4 - p_3 & 0 \\
e_3 = (v_2, v_5) & 0 & p_5 - p_2 \\
e_4 = (v_3, v_5) & 0 & p_5 - p_3 \\
\end{bNiceMatrix}
\]

This matrix is $4 \times 4$ if we consider 2D coordinate vectors, so the graph will be pinned rigid if the rank of this matrix is equal to 4. Interestingly, the pinned rigid graph can be represented as a rigid graph, as shown in the right figure in Fig. \ref{fig-pinned_ex} (b). The rigidity matrix $\mathcal{R}(G, p)$ for this representation can be written as:

\[
\begin{bNiceMatrix}[first-row, first-col]
\CodeBefore
\rectanglecolor{blue!15}{1-1}{3-3}
\rectanglecolor{green!15}{4-4}{7-5}
\Body
& v_1 & v_2 & v_3 & v_4 & v_5 \\
e_5 = (v_1, v_2) & p_1 - p_2 & p_2 - p_1 & 0 & 0 & 0 \\
e_6 = (v_1, v_3) & p_1 - p_3 & 0 & p_3 - p_1 & 0 & 0 \\
e_7 = (v_2, v_3) & 0 & p_2 - p_3 & p_3 - p_2 & 0 & 0 \\
e_1 = (v_1, v_4) & p_1 - p_4 & 0 & 0 & p_4 - p_1 & 0 \\
e_2 = (v_3, v_4) & 0 & 0 & p_3 - p_4 & p_4 - p_3 & 0 \\
e_3 = (v_2, v_5) & 0 & p_2 - p_5 & 0 & 0 & p_5 - p_2 \\
e_4 = (v_3, v_5) & 0 & 0 & p_3 - p_5 & 0 & p_5 - p_3 \\
\end{bNiceMatrix}
\]

Notice that the matrix has the form $\begin{bmatrix}
    A & 0 \\
    C & B
\end{bmatrix}$, where $A$ corresponds to the triangle (blue shaded part), $B$ is the pinned rigidity matrix part (green shaded part), and $C$ is the symmetric component from $B$. Since $\rank \begin{bmatrix}
    A & 0 \\
    C & B
\end{bmatrix} \geq \rank A + \rank B$ (see Appendix for details), the rank of $A$ is $S(3,2) = 3 \times 2 - 3$ (triangle), and $\rank B = 4$ (if it is pinned rigid). Therefore, the rank of the rigidity matrix of the right graph is $\geq 7$, and since $S(5,2) = 5 \times 2 - 3 = 7$, we can conclude that the right graph is rigid. Hence, if the graph is pinned rigid, then the converted graph will also be rigid as long as there are no algebraic relations between pinned points (for example, if $v_1 \sim v_3$ are on the same line, we would not have full rank of the rigidity matrix from the triangle). Please check Appendix (Lemma \ref{glulemma2}) for more details.

To summarize, we can illustrate the relationships between rigidity matroid, rigidity matrix, and pinned rigidity matrix using the following diagram:

\[
\begin{tikzcd}[row sep=large, column sep = large]
    G \arrow[d, "\text{matrix of }\mathbb{R}^d"] \arrow[r, "\text{positioning}"] & (G,p) \in \mathbb{R}^d \arrow[d] \arrow[r, "\text{pin points}"] & (\widetilde{G}, p) \arrow[d, ""] \\
    \mathcal{R}(G, x) \arrow[r, ""] & \mathcal{R}(G, p) \arrow[r, ""] & \mathcal{R}(\widetilde{G}, p) \\
\end{tikzcd}
\]

Notice that, since we assume all \(x_i\)s are distinct and due to the independence of \(x_i\)s, we obtain the following relations:
\[
S(n, d) \geq \rank \mathcal{R}(G, x) \geq \rank \mathcal{R}(G, p) \geq \rank \mathcal{R}(\widetilde{G}, p).
\]
Note that the rank of the pinned rigidity matrix is smaller than or equal to that of the rigidity matrix because it involves fewer vertices and edges.

Finally, we introduce a simplex, which is commonly used in topology.

\begin{definition}[\cite{Veljan2017TheDM}]\label{defsimplex}
    A $d$-dimensional simplex \(C = \{p_0, p_1, \dots, p_d\}\) is the smallest convex set (i.e., the convex hull) that contains \(d+1\) points \(p_0, p_1, \dots, p_d\) in \(d\)-dimensional space \(\mathbb{R}^d\) which are in general position (i.e., the vectors \(p_i-p_0\) form a basis in \(\mathbb{R}^d\)).
\end{definition}

\begin{figure}[H]
    \centering
    \begin{tikzpicture}[scale=0.4, transform shape]
        \node [fill=black,circle] (0) at (-4.75, 2.75) {};
        \node [fill=black,circle] (1) at (-1, 3.5) {};
        \node [fill=black,circle] (2) at (-2, 1) {};
        \node [fill=black,circle] (3) at (2.5, 3.75) {};
        \node [fill=black,circle] (4) at (1, 1.25) {};
        \node [fill=black,circle] (5) at (3.5, 1.25) {};
        \node [fill=black,circle] (6) at (8.25, 3.75) {};
        \node [fill=black,circle] (7) at (6.25, 1.5) {};
        \node [fill=black,circle] (8) at (9.5, 0.75) {};
        \node [fill=black,circle] (9) at (6.5, 4.25) {};
        \node [] (10) at (-4.75, -0.25) {};
        \node [] (11) at (-4.75, -0.25) {\LARGE $C_0$};
        \node [] (12) at (-1.5, -0.25) {};
        \node [] (13) at (-1.5, -0.25) {\LARGE $C_1$};
        \node [] (14) at (2.25, -0.25) {\LARGE $C_2$};
        \node [] (15) at (7.75, -0.25) {\LARGE $C_3$};
        \draw [-,thick] (1) to (2);
        \draw [-,thick] (3) to (4);
        \draw [-,thick] (4) to (5);
        \draw [-,thick] (5) to (3);
        \draw [-,thick] (9) to (7);
        \draw [-,thick] (9) to (8);
        \draw [-,thick] (9) to (6);
        \draw [-,thick] (6) to (8);
        \draw [-,thick] (6) to (7);
        \draw [-,thick] (7) to (8);
    \end{tikzpicture}
    \caption{{\it An illustration of $C_0, \dots, C_3$. $C_2$ is a triangle, and $C_3$ is a tetrahedron.}}
\end{figure}
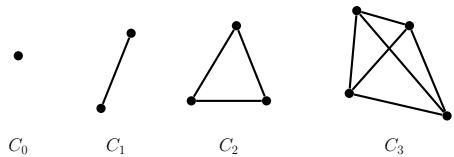

A simplex is a geometrical structure that represents a $d$-dimensional convex hull. It is unique (up to the uniqueness of the convex hull) and can be related to a rigid graph, which will be introduced later. Notice that a simplex can be considered a $d$-dimensional extension of a complete graph with no crossing edges and can be represented as a $K_{d+1}$ \cite{Veljan2017TheDM}. Since a simplex is both globally rigid (having only one realization) and infinitesimally rigid in $\mathbb{R}^d$, we can always take $C_{d-1}$ (where the number of vertices is the same as the dimension $d$) as pinned vertices and convert it to the pinned graph.

\section{Rigidity Condition for Graphs in $\mathbb{R}^d$}\label{sec03}

We will discuss the rigidity condition for a graph $G$ embedded in $\mathbb{R}^d$. We will focus primarily on the rigidity condition based on the union of two rigid graphs. Any rigid cluster can be constructed by the union of multiple rigid graphs in $\mathbb{R}^d$. In this section, we will examine the rigidity condition for graphs embedded in two-dimensional space for easier understanding, using the rigidity matrix, and in three-dimensional space by exploring examples in terms of rigid motion. We will define the general $d$-dimensional case in the next section.

\subsection{Rigidity Condition for Graphs in $\mathbb{R}^2$}
To determine whether a graph is rigid, one of the most well-known results is Laman's Theorem (1970), also known as the Geiringer-Laman Theorem \cite{Laman1970,Geiringer1927}. This theorem provides a criterion for rigidity in two dimensions and serves as a specific case of Theorem \ref{diso}, formulated without the use of a rigidity matrix. Cheng {\it et al.} extend this work by presenting conditions for rigidity in $\mathbb{R}^3$ using graph coverings \cite{CHENG2014} and Jord\'an {\it et al.} define a combinatorial characterization of rigid graphs in terms of vertex partitions and edge count conditions combined with spanning trees for $\mathbb{R}^d$ \cite{JORDAN2022v2}; however, to date, there is no equivalent version of Laman's Theorem in three dimensions without counterexamples. Cruickshank {\it et al.} \cite{cruickshank2024} establish conditions for global rigidity in $\mathbb{R}^d$ using the rigidity matrix, while Dewar {\it et al.} \cite{Dewar2021} explore rigidity in $l^d_p$. Related concepts include merging two graphs in 2D \cite{Changbin2006,Anderson2008} and decomposing graphs into rigid components \cite{Berg2003} within the context of developing Tay's theorem in the bar-joint framework. Lew {\it et al.} describe the connection of rigidity graphs in terms of eigenvalues of the stiffness matrix, which can be derived from the Laplacian \cite{lew2023}. Furthermore, the gluing lemma (Lemma \ref{glulemma}) shows that two rigid graphs joined by $d$ overlapping vertices remain rigid. However, prior studies examining the transition from Tay's Theorem (non-overlapping vertices) to the gluing lemma (with $d$ overlapping vertices) are limited. Therefore, this section will focus on the relationship between rigid graphs and methods for checking rigidity based on connections with $0$ to $d$ overlapping vertices.

\begin{theorem}[Geiringer-Laman Theorem \cite{Laman1970}\cite{Geiringer1927}\cite{wiki:gl}] A graph $G=(V,E)$ is generically rigid in $2$-dimensions with respect to bar-joint frameworks if and only if $G$ has a spanning subgraph $G'=(V,E')$ such that \begin{itemize} \item $|E'|=2|V|-3$, \item for all subsets $F\subset E'$, $|F|\leq 2|V(F)|-3$. \end{itemize} 
\end{theorem}

\begin{figure}
		\captionsetup{singlelinecheck = false, justification=raggedright}
		\begin{center}
\begin{tabular}{cc}
     \includegraphics[width=0.2\textwidth]{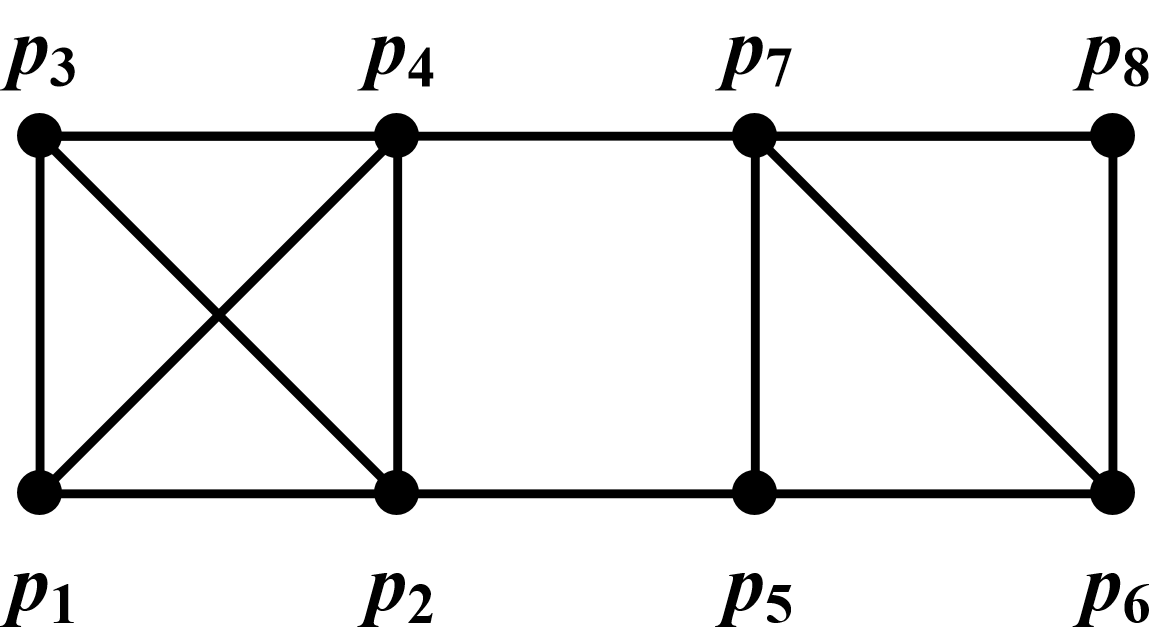}&
     \includegraphics[width=0.2\textwidth]{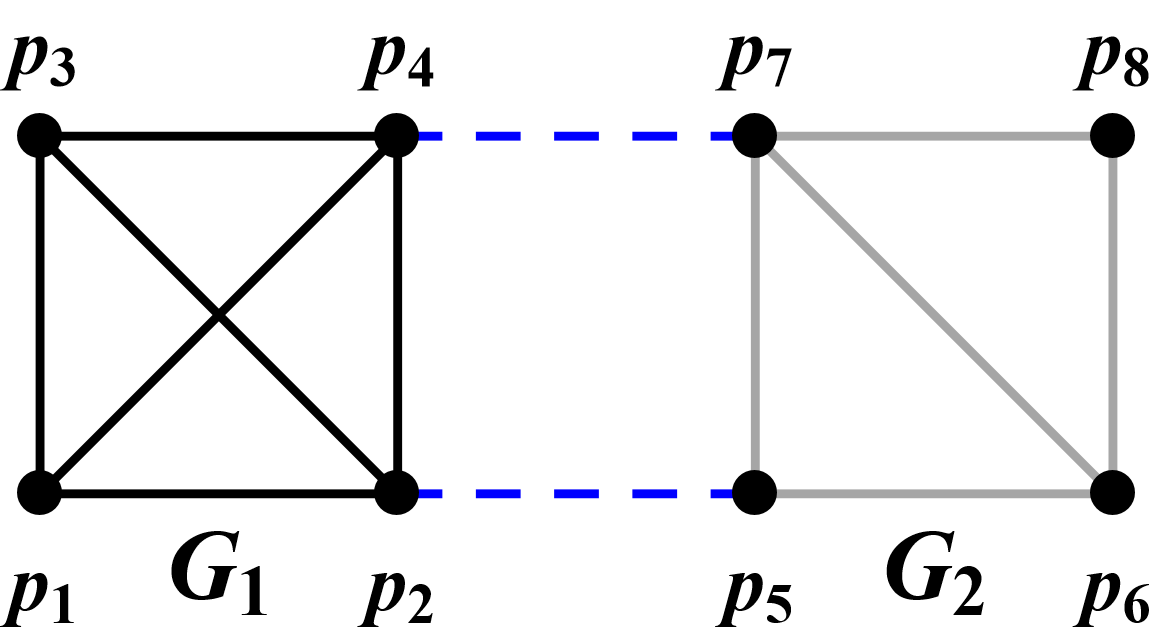}\\
     (a) & (b) \\ 
\end{tabular}

			\caption{{\it Counterexamples in Maxwell counting.} (a) A graph $G$ with locally over-constrained edges. (b) Segmentation of (a) into two rigid graphs.}
  \label{fig-ex-maxwell}
		\end{center}
	\end{figure}

A similar result for three-dimensional space was obtained in 1986 by Calladine and Pellegrino, often referred to as the Maxwell-Calladine Theorem. This theorem builds on Maxwell's conjecture (1864) \cite{Maxwell1864} and incorporates later contributions from Calladine \cite{CALLADINE1978,PELLEGRINO1986}. In granular systems, Maxwell counting, which forms the foundational principle of Laman's Theorem and is derived from Maxwell's conjecture, is the most commonly used method to describe the average contact number and properties of the system \cite{Lin2014,WN2020}. However, Maxwell counting is not always sufficient. For example, Fig. \ref{fig-ex-maxwell} presents a counterexample to Maxwell's counting method, demonstrating that simply checking the global condition of $|E| = 2|V| - 3$ is inadequate. Here, with 8 vertices, the required number of edges is $2 \times 8 - 3 = 13$, which matches the number of edges in graph $G$. However, this alone does not ensure the rigidity of the graph. If we use $p_1 = (0, 0)$, $p_2 = (1, 0)$, $p_3 = (0, 1)$, $p_4 = (1, 1)$, $p_5 = (2, 0)$, $p_6 = (3, 0)$, $p_7 = (2, 1)$, and $p_8 = (3, 1)$, we can calculate the rigidity matrix $\mathcal{R}(G, p)$ as
\begin{equation*}
\begin{bNiceMatrix}[first-row, first-col]
\CodeBefore
\rectanglecolor{blue!15}{1-1}{1-4}
\rectanglecolor{blue!15}{2-3}{2-6}
\rectanglecolor{blue!15}{3-5}{3-8}
\rectanglecolor{blue!15}{4-1}{4-2}
\rectanglecolor{blue!15}{4-7}{4-8}
\rectanglecolor{blue!15}{5-3}{5-4}
\rectanglecolor{blue!15}{5-7}{5-8}
\rectanglecolor{blue!15}{6-1}{6-2}
\rectanglecolor{blue!15}{6-5}{6-6}
\rectanglecolor{blue!15}{7-9}{7-12}
\rectanglecolor{blue!15}{8-11}{8-14}
\rectanglecolor{blue!15}{9-13}{9-16}
\rectanglecolor{blue!15}{10-9}{10-10}
\rectanglecolor{blue!15}{10-13}{10-14}
\rectanglecolor{blue!15}{11-11}{11-12}
\rectanglecolor{blue!15}{11-15}{11-16}
\rectanglecolor{blue!15}{12-7}{12-8}
\rectanglecolor{blue!15}{12-13}{12-14}
\rectanglecolor{blue!15}{13-3}{13-4}
\rectanglecolor{blue!15}{13-9}{13-10}
\Body  
&p_1& &p_2& &p_3& &p_4& &p_5& &p_6& &p_7& &p_8&\\
e_1&-1&0&1&0&0&0&0&0&0&0&0&0&0&0&0&0\\
e_2&0&0&1&-1&-1&1&0&0&0&0&0&0&0&0&0&0\\
e_3&0&0&0&0&-1&0&1&0&0&0&0&0&0&0&0&0\\
e_4&-1&-1&0&0&0&0&1&1&0&0&0&0&0&0&0&0\\        
e_5&0&0&0&-1&0&0&0&1&0&0&0&0&0&0&0&0\\
e_6&0&-1&0&0&0&1&0&0&0&0&0&0&0&0&0&0\\
e_7&0&0&0&0&0&0&0&0&-1&0&1&0&0&0&0&0\\
e_8&0&0&0&0&0&0&0&0&0&0&1&-1&-1&1&0&0\\
e_{9}&0&0&0&0&0&0&0&0&0&0&0&0&-1&0&1&0\\
e_{10}&0&0&0&0&0&0&0&0&0&-1&0&0&0&1&0&0\\
e_{11}&0&0&0&0&0&0&0&0&0&0&0&-1&0&0&0&1\\
e_{12}&0&0&0&0&0&0&-1&0&0&0&0&0&1&0&0&0\\
e_{13}&0&0&-1&0&0&0&0&0&1&0&0&0&0&0&0&0
\end{bNiceMatrix}
\end{equation*}

Each row corresponds to an edge, and the blue shaded block represents vertices connected to that edge. For example, the first row is for $e_1$, and $e_1$ is connected to $p_1$ and $p_2$; thus, the value corresponding to $(p_1 - p_2)$ is placed at $p_1$, and $-(p_1 - p_2)$ is placed at $p_2$. By calculating the rigidity matrix $\mathcal{R}(G,p)$, we observe that this graph is not rigid because $\mathcal{R}(G,p)$ has a rank of 12. The graph contains a square $(p_2, p_4, p_7, p_5)$ which is not rigid along the line $p_4 - p_5$. This is because $(p_2, p_3)$ is an excess edge and could be eliminated (or $(p_1, p_4)$). Moving the edge $(p_2, p_3)$ to $(p_4, p_5)$ would increase the rank of the rigidity matrix to 13. Instead of relocating edges, consider segmenting the graph $G$ into two rigid graphs: Fig. \ref{fig-ex-maxwell} (b) shows $G_1$ (black-colored, $p_1, p_2, p_3, p_4$) and $G_2$ (gray-colored, $p_5, p_6, p_7, p_8$), connected via two lines $(p_2, p_5)$ and $(p_4, p_7)$. Since both graphs are constructed from multiple triangles, they are rigid. Joining two rigid graphs with no overlapping vertices requires at least three edges to maintain rigidity; adding only two edges would create a tetragon, which is not rigid.

\begin{figure}
		\captionsetup{singlelinecheck = false, justification=raggedright}
		\begin{center}
\begin{tabular}{cc}
     \includegraphics[width=0.25\textwidth]{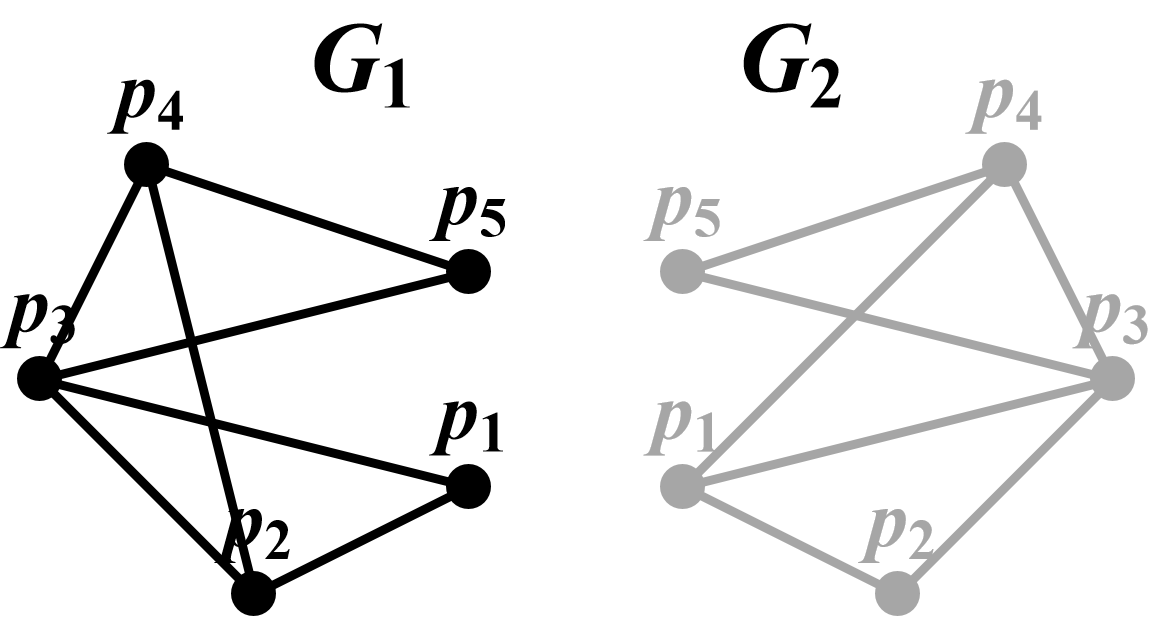}&
     \includegraphics[width=0.2\textwidth]{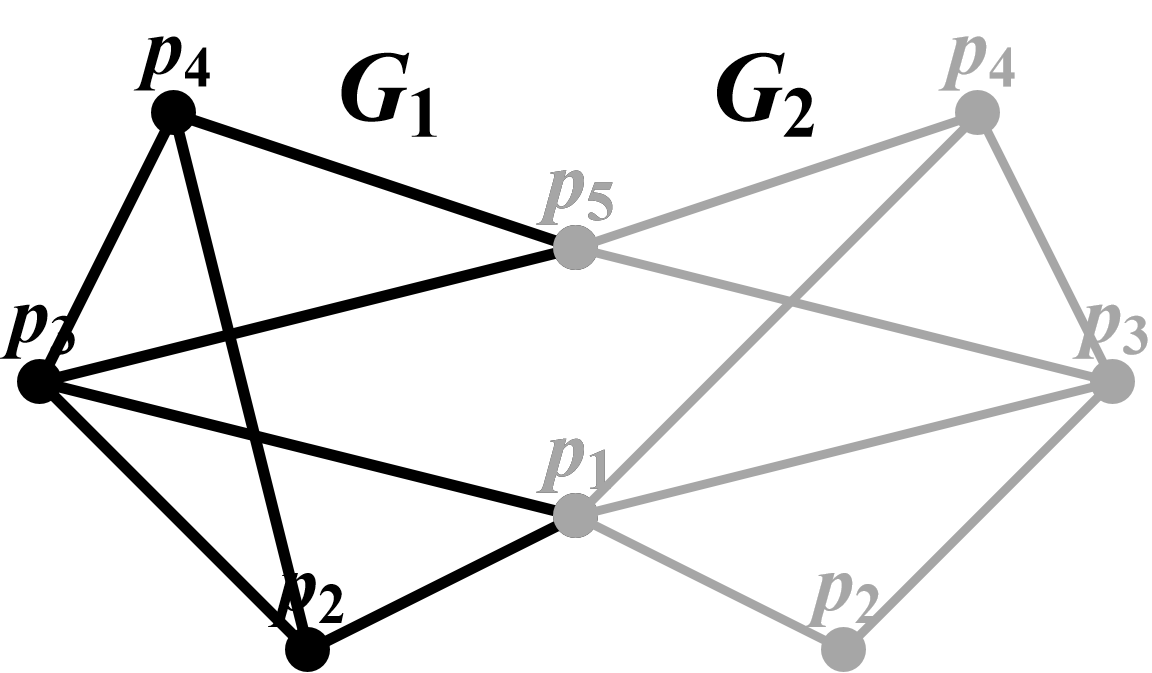}\\
     (a) & (b) \\ 
     \includegraphics[width=0.2\textwidth]{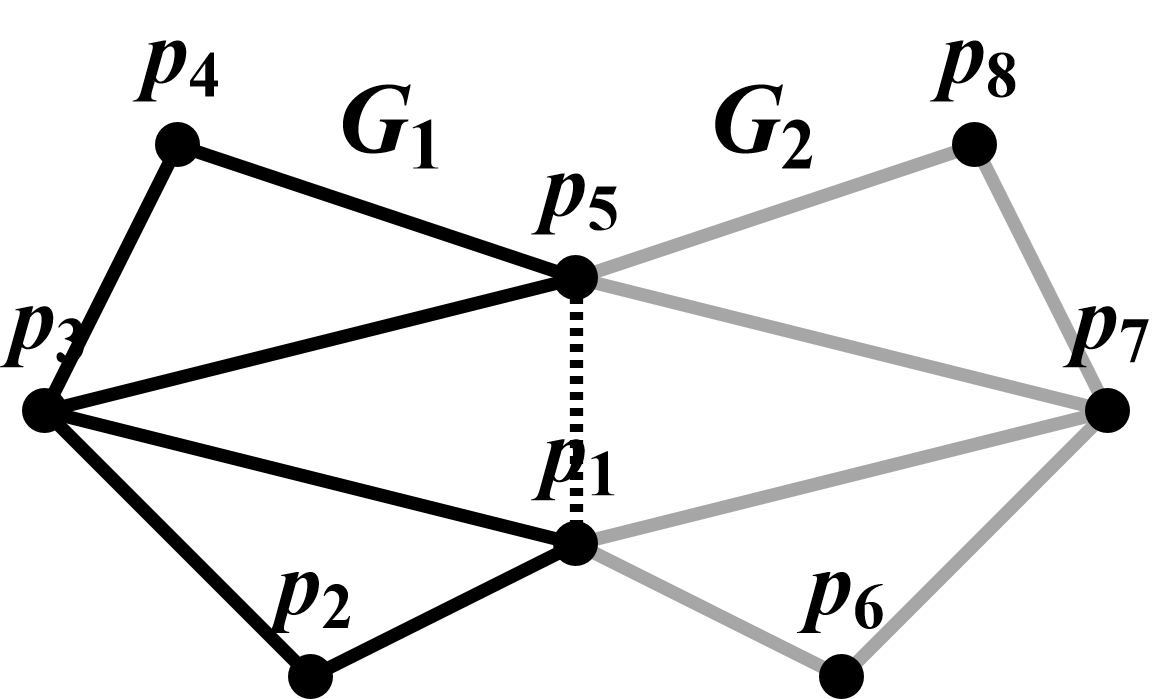}&
     \includegraphics[width=0.2\textwidth]{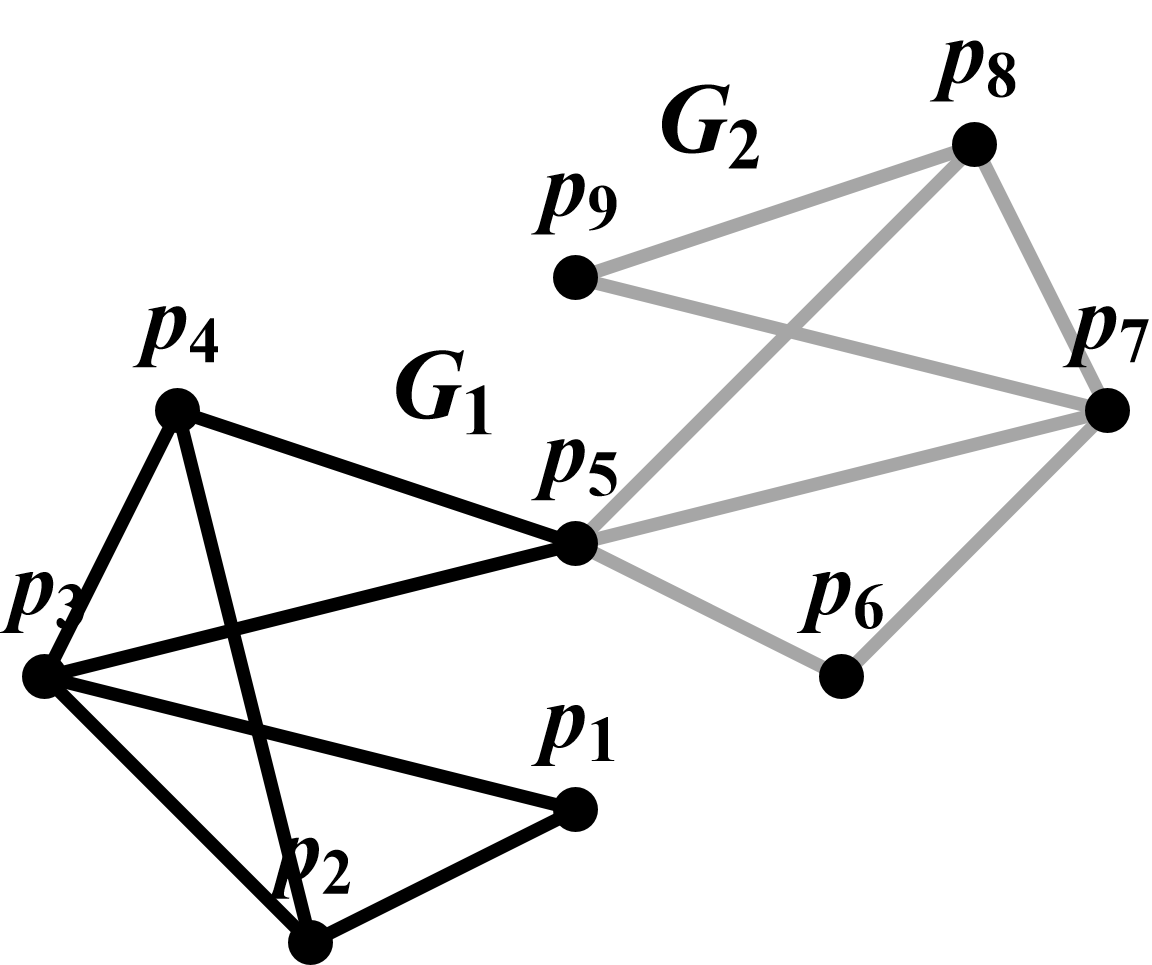}\\
     (c) & (d) \\ 
\end{tabular}

			\caption{{\it Combining two rigid graphs.} (a) Two graphs, $G_1$ and $G_2$. (b) $G_1$ and $G_2$ joined via two overlapping vertices. (c) The graph from (b) with one additional edge. (d) $G_1$ and $G_2$ joined via one overlapping vertex.}
  \label{fig-ex-2d}
		\end{center}
	\end{figure}

Next, we will consider graphs joined with different numbers of vertices and the conditions required to maintain rigidity. Fig. \ref{fig-ex-2d} (a) shows two rigid graphs: $G_1$ on the left (black-colored graph) and $G_2$ on the right (gray-colored graph), each with a rigidity matrix rank of $7$. Fig. \ref{fig-ex-2d} (b) represents a graph formed by connecting $G_1$ and $G_2$ via $p_1$ and $p_5$. The rank of the rigidity matrix for this graph $G_1 \cup G_2$ is $13$, indicating that it is rigid, although there is an excess edge that could be removed.

An interesting feature is that, since the number of overlapping vertices is $2$ in two-dimensional space, we can consider the gray graph in Fig. \ref{fig-ex-2d} (b) as {\it pinned rigid}. The rank of the pinned rigidity matrix $\mathcal{R}(\widetilde{G_2},p)$ becomes $6$ (as $p_1$ and $p_5$ are not included in the pinned rigidity matrix). Using Lemma \ref{glulemma2}, the rank of the rigidity matrix $\mathcal{R}(G_1 \cup G_2,p)=\mathcal{R}(G_1 \cup \widetilde{G_2},p)$ is $7 + 6 = 13$.

To create a minimally rigid graph, we can rearrange an edge $(p_2, p_4)$ of $G_1$ to $(p_1, p_5)$ and an edge $(p_1, p_4)$ of $G_2$ to $(p_1, p_5)$, joining them with two overlapping vertices and one edge connected to them, as shown in Fig. \ref{fig-ex-2d} (c). The dashed line $(p_1, p_5)$ represents the shared edge between $G_1$ and $G_2$.

Next, we will consider the case where there is only one overlapping vertex. If we join $G_1$ and $G_2$ via $p_5$, $G_2$ is no longer pinned rigid because $1 \leq d = 2$. The graph $G_2$ can rotate, so we need at least one additional edge to prevent rotation. In a generic condition, we can add any edge that connects $G_1$ and $G_2$. However, in Fig. \ref{fig-ex-2d} (d), since $(p_1, p_5, p_9)$ are collinear, we need to add an edge that is not algebraically dependent (e.g., an edge $(p_1, p_8)$). Alternatively, we can ensure generic conditions by assigning different $x$ coordinates to $p_1$, $p_5$, and $p_9$ so they do not lie on the same line. For simplicity, we will assume that all graphs have algebraically independent edges.
To summarize, when joining two rigid graphs $G_1$ and $G_2$ embedded in $\mathbb{R}^2$, we need:

\begin{itemize} \item At least $3$ additional edges between $G_1$ and $G_2$ to maintain rigidity if $|V(G_1 \cap G_2)| = \emptyset$. These additional edges should not form any partially rigid graphs when attached to $G_1$ or $G_2$. \item At least $1$ additional edge between $G_1$ and $G_2$ to maintain rigidity if $|V(G_1 \cap G_2)| = 1$. \item No additional edge is required if $|V(G_1 \cap G_2)| = 2$. \item *If $|V(G_1 \cap G_2)| = \emptyset$ and any of these additional edges form rigid graphs between $G_1$ and $G_2$, then $G_1$ and $G_2$ need to be updated to $G_1'$ and $G_2'$ before the above rules can be applied. \end{itemize}

We will further elaborate on the first condition. For instance, in Fig. \ref{fig-ex-maxwell}(b), adding edges $(p_2, p_5)$, $(p_4, p_5)$, and $(p_4, p_7)$ creates two new triangles. In this case, the graph becomes rigid but satisfies the first condition rather than the third. This scenario aligns with concepts introduced by Olfati-Saber and Murray, where these types of connections—$(p_2, p_5)$, $(p_4, p_5)$, and $(p_4, p_7)$—are referred to as Z-links (also represented by the dashed network in Fig. \ref{fig-body-bar}(b)). Z-links are used to connect disjoint networks while preserving network rigidity \cite{OS2002,OS2002Rep}. The Z-type linkage, which involves constructing simplices, is fundamental for applying Tay's theorem in bar-joint rigidity and will be examined more closely in three-dimensional space in the following section.

Moreover, these new edges must connect a vertex in $G_1$ to a vertex in $G_2$, an idea similar to $d$-dimensional algebraic connectivity in terms of rigid partition discussed in \cite{lew2023}. These edges also need to be independent of the existing edges. If additional edges, such as $(p_3, p_7)$ or $(p_4, p_8)$, provide equivalent connections to $(p_4, p_7)$, the graph would lose its rigidity.

\subsection{Rigidity condition of Graph in $\mathbb{R}^3$}
\begin{figure}
		\captionsetup{singlelinecheck = false, justification=raggedright}
		\begin{center}
    \includegraphics[width=0.5\textwidth]{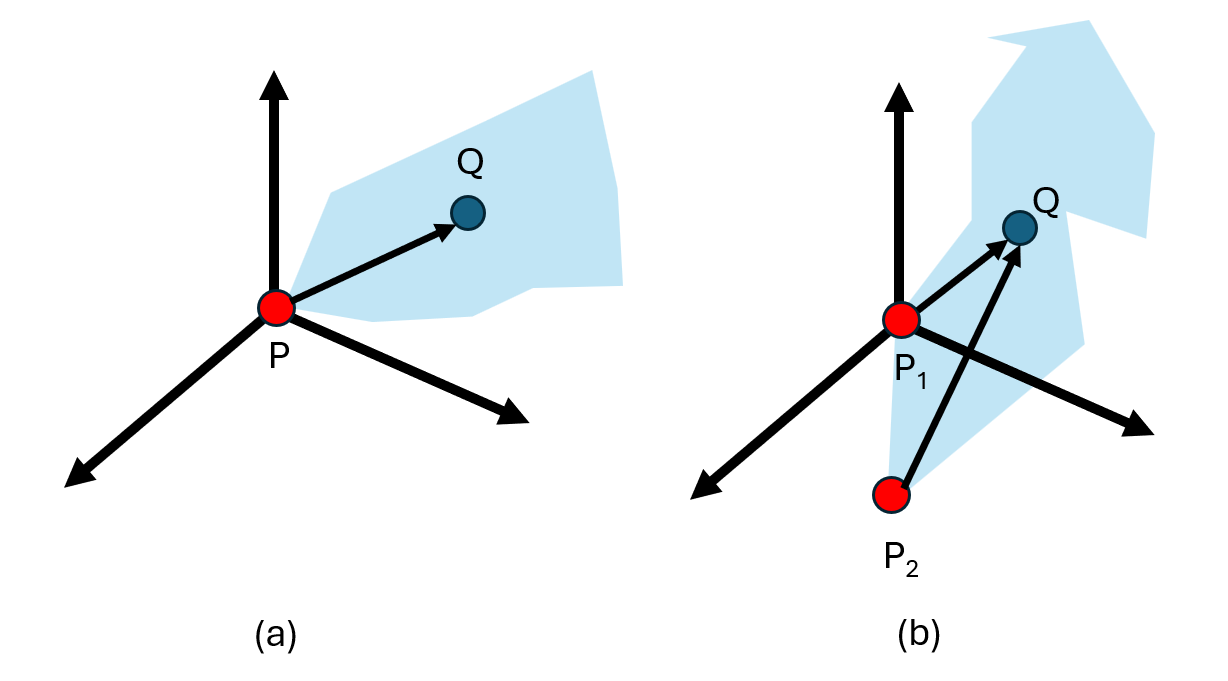}
			\caption{{\it Graphical interpretation of rigid body motions in 3D with pinned points.} (a) A rigid graph fixed via one, red pinned point. (b) A rigid graph fixed via two, red pinned points.}
  \label{fig-pinned-3d}
		\end{center}	
	\end{figure}

In this section, we will explore the three-dimensional case. First, we examine the scenario of a pinned graph with $|P| < d = 3$ in the context of rigid body motion. In three dimensions, there are $6$ possible rigid body motions. Consider the case with one pinned vertex, as illustrated in Fig. \ref{fig-pinned-3d} (a). This scenario is analogous to determining the location of a point $Q$ (e.g., the center of mass of a rigid object) relative to a base point $P$ for a rigid body. The distance between $P$ and $Q$, represented by an arrow in Fig. \ref{fig-pinned-3d}, indicates that the rigid graph, which does not deform continuously, can be treated as a rigid body. The center of mass $Q$ connected to $P$ by a rigid bar (the arrow in the figure can rotate within the $(xy)$, $(yx)$, and $(zx)$ planes. Consequently, the sphere equation describing the locus of $Q$ based on the given distance will have three unknowns, corresponding to three degrees of freedom. To fix $Q$ in space, at least three edges connecting to other pinned points are required.

\begin{figure}
		\captionsetup{singlelinecheck = false, justification=raggedright}
		\begin{center}
\begin{tabular}{cc}
     \includegraphics[width=0.25\textwidth]{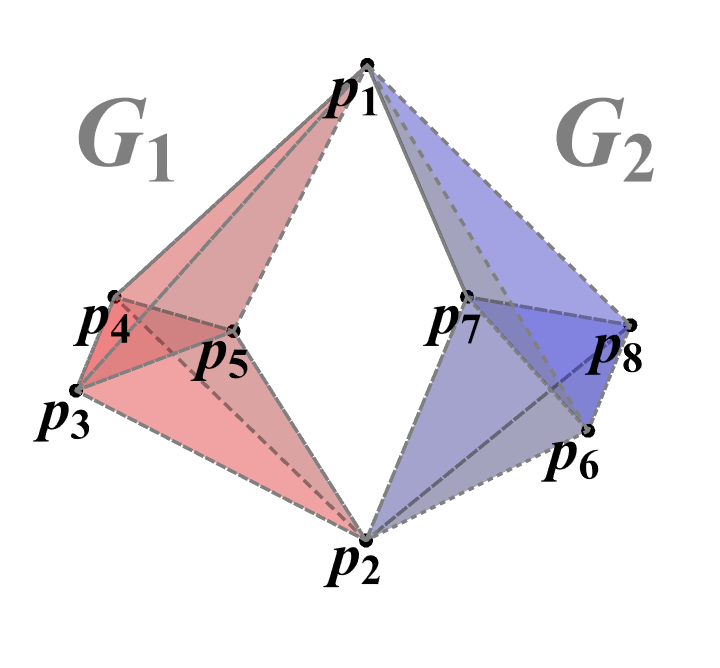}&
     \includegraphics[width=0.25\textwidth]{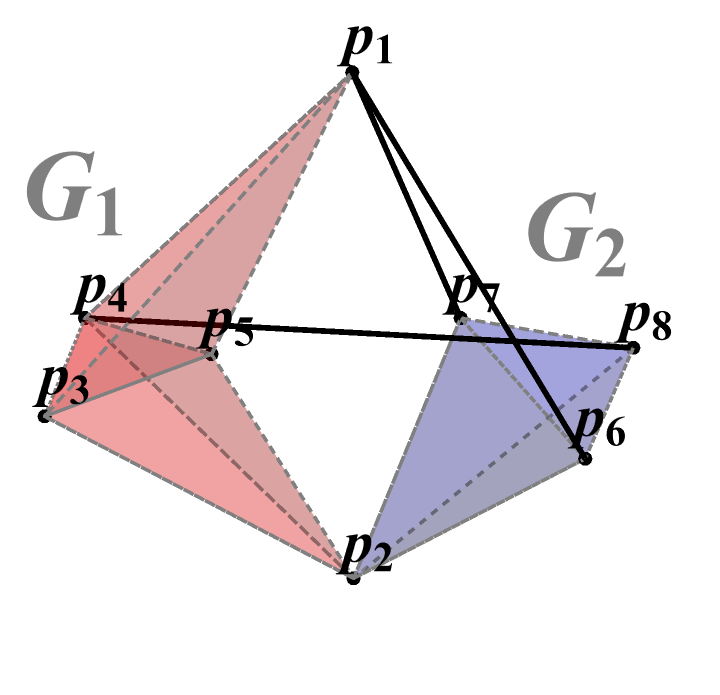}\\
     (a) & (b) \\ 
     \includegraphics[width=0.25\textwidth]{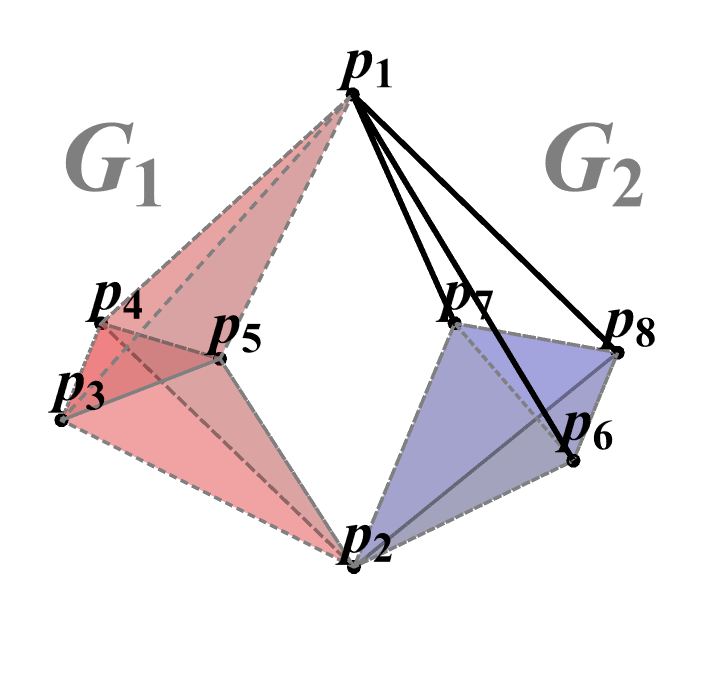}&
     \includegraphics[width=0.25\textwidth]{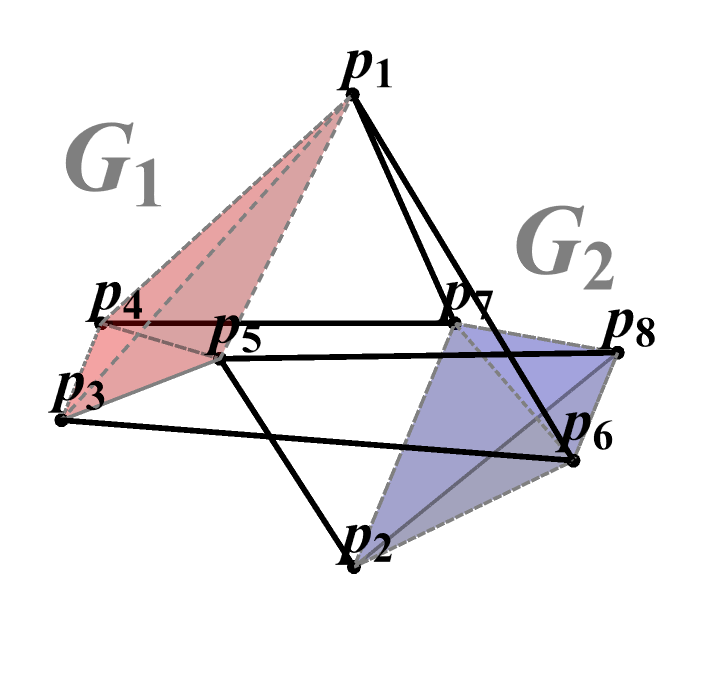}\\
     (c) & (d) \\ 
\end{tabular}

			\caption{{\it Combining two rigid graphs in $\mathbb{R}^3$.} (a) Double banana graph composed of $G_1$ and $G_2$. (b) Combination of $G_1$ and $G_2$ with one overlapping vertex and proper edge connection. (c) Combination of $G_1$ and $G_2$ with one overlapping vertex and improper edge connection. (d) $G_1$ and $G_2$ joining with no overlapping vertices.}
  \label{fig-ex-3d}
		\end{center}
	\end{figure}

In the second case, shown in Fig. \ref{fig-pinned-3d} (b), where $|P| = 2$, the vertices $P_1$ and $P_2$ form a triangle with $Q$. Although a triangle itself does not change its shape, it can rotate around the $z$-axis because $P_1$ and $P_2$ are pinned on the $xy$-plane. Therefore, one additional edge is necessary to fix $Q$ and prevent this rotation.
To understand this concept more clearly, consider the example shown in Fig. \ref{fig-ex-3d} consisting of a double banana graph and some of its variations, which serve as a counterexample to Laman's Theorem in three dimensions~\cite{Thorpe2002}. Specifically, in Fig. \ref{fig-ex-3d} (a), each vertex is connected to at least four edges, which is a stronger condition than the 3-connected requirement. Furthermore, the graph has 18 edges, which corresponds to the formula $8 \times 3 - 6 = 18$, satisfying the counting condition. Despite this, the graph is not rigid, as $G_1$ and $G_2$ can move along the line $(p_1, p_2)$.

Upon closer inspection, $G_1$ (shaded red) and $G_2$ (shaded blue) are individually rigid. Based on the previous discussion on rigid body motion, we conclude that, since $G_1$ and $G_2$ are joined via two vertices, one additional edge \textbf{crossing between $G_1$ and $G_2$} is required to make $G_1 \cup G_2$ rigid. Given that the optimal number of edges for a minimally rigid graph with 8 vertices is 18, this implies that the graph $G$ is over-constrained. Specifically, since there is no edge between $p_1$ and $p_2$, no edge will be removed when joining the two graphs, similar to Fig. \ref{fig-ex-2d} (c). Therefore, by adding a \textbf{crossing edge} between $G_1$ and $G_2$, we can achieve a minimally rigid graph by removing one edge, as shown in Fig. \ref{fig-ex-3d} (b).

However, if a different edge is added instead of the crossing edge, as depicted in Fig. \ref{fig-ex-3d} (c), the result will be the same as the original graph (a) and will not be rigid. The reason is that the new edges $(p_1, p_6)$, $(p_1, p_7)$, and $(p_1, p_8)$ form a tetrahedron, and combining these with $G_2$ generates a larger rigid cluster composed of two tetrahedrons. This increases the number of joints between the two rigid bodies to two instead of one, and there will be no \textbf{crossing edge} between these rigid graphs. Thus, the new graph $G_1 \cup G_2$ will not be rigid. Therefore, additional edges should not form any rigid structure on their own, as this would alter the number of joints connecting the rigid bodies.

In summary, when joining two graphs without overlapping vertices, six edges are required to ensure rigidity, corresponding to the three transitional $(x, y, z)$ and three rotational $(xy, yz, zx)$ degrees of freedom. Additionally, these six edges should not form any tetrahedral structures, as illustrated in Fig. \ref{fig-ex-3d} (d). Although there is one triangle formed by $(p_1, p_6)$ and $(p_1, p_7)$ combined with $G_2$, there is no tetrahedron among the additional edges. This aligns with the requirement of nested triangles in three dimensions, as discussed in \cite{BRUUN2022}, where tetrahedrons were joined based on planar views of nested triangles and the necessary connections and supports were estimated.

Thus, when joining two rigid graphs $G_1$ and $G_2$ embedded in $\mathbb{R}^3$, the following conditions are necessary to maintain rigidity:

\begin{itemize} 
\item If $|V(G_1 \cap G_2)| = \emptyset$, at least $6$ additional edges need to be added between $G_1$ and $G_2$ to ensure rigidity. These additional edges should not form any partially rigid structures when attached to $G_1$ or $G_2$. 
\item If $|V(G_1 \cap G_2)| = 1$, at least $3$ additional edges are required between $G_1$ and $G_2$ to maintain rigidity. \item If $|V(G_1 \cap G_2)| = 2$, at least $1$ additional edge is needed between $G_1$ and $G_2$ to ensure rigidity. 
\item No additional edge is required if $|V(G_1 \cap G_2)| = 3$. 
\item *If $|V(G_1 \cap G_2)| = \emptyset$ and any of the additional edges form rigid structures between $G_1$ and $G_2$, then $G_1$ and $G_2$ need to be updated to $G_1'$ and $G_2'$ before applying the above rules. 
\end{itemize}

Note that these additional edges should connect one vertex in $G_1$ to another vertex in $G_2$, similar to the two-dimensional case.
\subsection{General $d$-dimensional Space}
So far, we have discussed rigidity conditions for 2D and 3D graphs. In this section, we generalize the concept to rigidity in $d$-dimensional space. Again, we consider a generic framework where there are no algebraically dependent edges (e.g., no two distinct edges are collinear). In $d$-dimensional space, there are $d$ translational degrees of freedom (each representing linear motion along a coordinate axis) and $\frac{d(d-1)}{2}$ rotational degrees of freedom (corresponding to the number of planes perpendicular to the rotation axis). To generalize this concept, we define $r$-dimensional motion as follows:

\begin{definition} An $r$-dimensional motion is defined as a homogeneous motion in $r$ dimensions, based on a choice of $r$ basis vectors from the $d$-dimensional coordinate basis $(x_1, \dots, x_d)$ in $\mathbb{R}^d$. Specifically, for the 1- and 2-dimensional cases, the motion can be written in the form $c_0{x_0}^r + c_1{x_1}^r + \dots + c_r{x_r}^r$ (up to affine deformation in $\mathbb{R}^d$), where $x_0$ is an arbitrary constant, and $c_i$ is either 0 or 1, depending on the choice of $r$ basis vectors from the $d$ available dimensions. \end{definition}

To count the number of $r$-dimensional motions, we use the binomial coefficient $\frac{d!}{r!(d-r)!}$ to account for the possible combinations. For $r > 2$, we approximate higher-dimensional motions as combinations of translational and rotational motions. Since a graph is a linearly connected object, high-dimensional motions are typically not considered. However, if one needs to approximate objects in $d$-dimensional space, it may be necessary to consider whether higher-dimensional motions (for $r \geq 3$) should be included. Thus, the total number of degrees of freedom is $d + \frac{d(d-1)}{2}$, which simplifies to $\frac{d(d+1)}{2} = \binom{d+1}{2}$. This quantity corresponds to the difference between the total number of columns in the rigidity matrix, $d|V|$, and its full rank, $d|V| - \binom{d+1}{2}$.

Before presenting the main theorem, we need to define the concept of a maximal rigid cluster within a graph. As illustrated with the double banana example in $\mathbb{R}^3$, the maximal condition is crucial for maintaining rigidity when combining different rigid graphs.

\begin{definition} The maximal rigid clusters of a graph $G$, denoted as $\max(G)$, is defined as the set of the largest rigid clusters that $G$ can have. This implies that no combination of extra edges that do not belong to $\max(G)$ can form rigid graphs (or $d$-dimensional simplices) partially connected to $\max(G)$. 
\end{definition}
For example, in Fig. \ref{fig-ex-maxwell} (a), there are two rigid graphs, so $\max(G) = {G_1, G_2}$. For an arbitrary graph $G$, $\max(G)$ may not be unique. However, for simplicity, we will consider the case where $G$ is a rigid graph and $\max(G) = {G}$. We will now generalize the rules for gluing two rigid graphs while preserving rigidity, as explored in the previous section.

Let $G_A$ be a rigid graph defined by $G_A = G(V_A, E_A)$ and $G_B$ be a rigid graph defined by $G_B = G(V_B, E_B)$, where $|V_A| = a$ and $|V_B| = b$. The rank of the rigidity matrix for $G_A$ is $\rank \mathcal{R}(G_A,p) = da - \binom{d+1}{2}$, and for $G_B$, it is $\rank \mathcal{R}(G_B,p) = db - \binom{d+1}{2}$.

\begin{theorem} Suppose the new graph $G = G_A \cup G_B$ is formed by overlapping $m$ vertices, where $|V(G_A \cap G_B)| = m$ and $m < d$. To ensure that $G$ is infinitesimally rigid or that its rigidity matrix has rank $\rank \mathcal{R}(G,p) = d|V(G)| - \binom{d+1}{2}$, we need at least $\binom{d - m + 1}{2}$ additional edges $E'$ between $G_A$ and $G_B$. If $m \geq d$, then the number of additional edges required is given by $[\binom{d - m + 1}{2}]$, where $[X]$ denotes the function that returns $X$ if $X \in \mathbb{Z}^+$, and $0$ otherwise. \end{theorem}

We will prove this by considering different cases for $m$: \begin{itemize} \item Case 1: $m = 0$, \item Case 2: $m \geq d$, \item Case 3: $0 < m < d$. \end{itemize}

\begin{proof} 
\textbf{Case 1: $m = 0$.}
Suppose $G_A$ and $G_B$ are two rigid graphs with no overlapping vertices ($m = 0$). Since $G_A \cap G_B = \emptyset$, the graph $G_A \cup G_B$ is given by the direct sum of the graphs $G_A$ and $G_B$. Therefore by Lemma \ref{matrank2}, we have: 
$\rank \mathcal{R}(G_A \cup G_B,p)=\rank \mathcal{R}(G_A \oplus G_B,p)= \rank \mathcal{R}(G_A,p) + \rank \mathcal{R}(G_B,p)$.

According to Lemmas \ref{Srankmatrix} and \ref{diso}, the ranks of the rigidity matrices for $G_A$ and $G_B$ are: $\rank \mathcal{R}(G_A,p) = da - \binom{d+1}{2}$ and $\rank \mathcal{R}(G_B,p) = db - \binom{d+1}{2},$
respectively. Thus, the rank of the rigidity matrix for $G_A \cup G_B$ is:
\begin{align*}
\rank \mathcal{R}(G_A \cup G_B,p) &= (da - \binom{d+1}{2}) + (db - \binom{d+1}{2}) \\
&= d(a + b) - 2 \binom{d+1}{2}.
\end{align*}

Let $|V(G_A \cup G_B)|$ denote the number of vertices in $G_A \cup G_B$. The required rank for $G_A \cup G_B$ to satisfy the rigidity condition is:
\[
d|V(G_A \cup G_B)| - \binom{d+1}{2}=d(a+b)- \binom{d+1}{2}.
\]
Thus, to meet this requirement, we need to add: $\binom{d+1}{2}$ additional edges.
\end{proof}

This result is analogous to Tay's theorem (1984) \cite{TAY1984} for body-bar rigidity, which requires $\binom{d+1}{2}$ additional edges (or rigid linkages) to satisfy the rigidity condition. In this context, the rigid body does not need to be globally rigid, but it must satisfy the maximally rigid graph condition.
\begin{proof} 
\textbf{Case 2: $m \geq d$.}

If $m \geq d$, by Lemma \ref{glulemma2}, $G_B$ can be converted to $G(I_B, P_B, E_B)$ with $|P_B| = m$, which is pinned rigid. Hence, no additional edges are required to ensure rigidity. Therefore, $[\binom{d - m + 1}{2}] = 0$.
\end{proof}
\textbf{Case 3: $0 < m < d$.}

For this case, we consider the following scenarios: 
\begin{itemize} 
\item Subcase 1: $G_A$ and $G_B$ have the same overlapped edges with $|E(G_A \cap G_B)| = \binom{m}{2}$, 
\item Subcase 2: $G_A$ and $G_B$ have the same overlapped edges with $|E(G_A \cap G_B)| < \binom{m}{2}$, 
\item Subcase 3: $G_A$ and $G_B$ have different overlapped edges with $|E(G_A \cap G_B)| \leq \binom{m}{2}$. 
\end{itemize}

Since the proofs for Subcases 2 and 3 can be derived from Subcase 1, we will focus on Subcase 1.

\begin{proof}
\textbf{Subcase 1: $0 < m < d$ and $G_A$ and $G_B$ have the same overlapped edges with $|E(G_A \cap G_B)| = \binom{m}{2}$.}

For a subgraph $G_A \cap G_B$ with $|V(G_A \cap G_B)| = m$, to maintain rigidity, this subgraph must be a complete graph on $m$ vertices, thus having $\binom{m}{2}$ edges. When merging $G_A$ and $G_B$, we lose $\binom{m}{2}$ edges because they are counted twice in the union.

The rigidity matrix rank condition requires:
\[
\rank \mathcal{R}(G_A \cup G_B) = d(a + b - m) - \binom{d+1}{2}.
\]

To be minimally rigid, we need:
\[
|E(G_A \cup G_B)| = d(a + b - m) - \binom{d+1}{2}.
\]

We can express this in terms of the rigidity matrices of $G_A$ and $G_B$:
\begin{align*}
&\rank \mathcal{R}(G_A \cup G_B,p) \\
= &\rank \mathcal{R}(G_A,p) + \rank \mathcal{R}(G_B,p) - \binom{m}{2} + x \\
&= da - \binom{d+1}{2} + db - \binom{d+1}{2} - \binom{m}{2} + x.
\end{align*}

Solving for $x$ (the number of additional edges needed):
\begin{align*}
&d(a + b - m) - \binom{d+1}{2} \\
= &da - \binom{d+1}{2} + db - \binom{d+1}{2} - \binom{m}{2} + x \\
x &= d(a + b - m) - da - db + \binom{d+1}{2} + \binom{m}{2} \\
&= -dm + \binom{d+1}{2} + \binom{m}{2}.
\end{align*}

Simplifying:
\begin{align*}
x &= \frac{d^2 - 2dm + m^2 + d - m}{2} \\
&= \frac{(d - m)(d - m + 1)}{2} \\
&= \binom{d - m + 1}{2}.
\end{align*}

Thus, at least $\binom{d - m + 1}{2}$ additional edges are required.
\end{proof}

\textbf{Subcase 2 and Subcase 3:}

For the other cases, the proof follows similarly by considering the number of overlapping edges and adjusting for the specific conditions. The argument shows that regardless of how the edges are connected or rearranged, the number of additional edges required to maintain rigidity remains consistent with the result obtained in Subcase 1.

Thus, regardless of the connection, we need the same number of supports (edges) to maintain rigidity. This can also be explained through the rigid body motion based on $m$ number of joints, which generates additional degrees of freedom due to $d-m$ dimensional translation and rotation motions. Hence, we get $\binom{d - m + 1}{2}$ additional edges.
\section{Algorithm to check rigidity of Graph in $\mathbb{R}^d$}\label{sec04}
Now, how can we check whether the graph $G$ is rigid? We will first show how to construct rigid graphs. 
\subsection{Method to construct rigid graph in $\mathbb{R}^d$}
In this section, we will address different ways (or steps) to construct rigid graphs. First, we need to define a subgraph structure $I_d$.
\begin{definition}
    $I_d$ is composed with one vertex plus $d$ number of algebraically independent edges of each other (i.e., vectors form $n$ number of bases) such that the pinned rigid matrix $\mathcal{R}(I_d,p)$ of the size of $d\times d$ has a full rank.
\end{definition}
This $I_d$ is related to Henneberg sequences (a method to construct a rigid graph) type I in $d$-dimensional space which we will introduce next.
\begin{figure}[H]
    \centering
\begin{tikzpicture}[scale=0.50, transform shape]
		\node [fill=black,circle] (0) at (-5.5, 3.25) {};
		\node [] (1) at (-7, 0.5) {};
		\node [] (2) at (-5, 0.25) {};
		\node [fill=black,circle] (3) at (-1, 3.25) {};
		\node [] (4) at (-2.5, 0.5) {};
		\node [] (5) at (-0.5, 0.25) {};
		\node [] (6) at (0.5, 2) {};
		\node [fill=black,circle] (7) at (4.25, 3.25) {};
		\node [] (8) at (2.75, 0.5) {};
		\node [] (9) at (4.75, 0.25) {};
		\node [] (10) at (5.75, 2) {};
		\node [] (11) at (2.5, 2.25) {};
		\node [] (12) at (-6, -0.75) {\LARGE $I_2$};
		\node [] (13) at (-1, -0.75) {\LARGE $I_3$};
		\node [] (14) at (4, -0.75) {\LARGE $I_4$};
		\draw [-,thick](0) to (1.center);
		\draw [-,thick](0) to (2.center);
		\draw [-,thick](3) to (4.center);
		\draw [-,thick](3) to (5.center);
		\draw [-,thick](3) to (6.center);
		\draw [-,thick](7) to (8.center);
		\draw [-,thick](7) to (9.center);
		\draw [-,thick](7) to (10.center);
		\draw [-,thick](11.center) to (7);
\end{tikzpicture}
    \caption{{\it An illustration of $I_d$ for $n=2,3,4$.} $I_3$ can be thought of as a tripod in three-dimensional space.}
\end{figure}
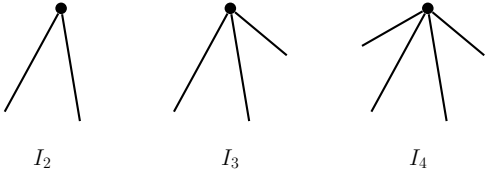
The Henneberg sequences are designed to generate minimally rigid graphs in two-dimensional space (called Laman graphs) and minimally rigid graphs in three-dimensional space (called Geiringer graphs).
\begin{definition}[{\cite{Borcea04}}\cite{Capco18}\cite{Grasegger18}]
	Henneberg sequences are used to construct Laman graph inductively such that, for a graph $G$, there is a sequence $G_3, G_4,\dots, G_n$ of Laman graphs on $3,4,\dots,n$ vertices. $G_3$ is a triangle, $G_n=G$ and each graph $G_{i+1}$ is obtained from the previous one $G_i$ via one of two types of steps
	\begin{enumerate}
		\item Type I $\colon$ adds a new vertex and two new edges connecting this vertex to two arbitrary vertices of $G_i$.
		\item Type II $\colon$ adds a new vertex and three new edges, and removes an old edge. The three new edges must connect the new vertex to three old vertices, such that at least two of them are joined via an edge which will be removed.
	\end{enumerate}
\end{definition}
Fig. \ref{fig-Henneberg-2d} shows Henneberg sequence (b) type I and (c) type II.
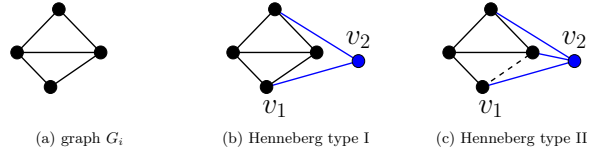
\begin{figure}[h]
	\begin{center}
		
		\begin{tikzpicture}[scale=0.65, transform shape]
			\begin{groupplot}[group style={group name=ch1plots, group size=3 by 1},xmin=-1.5,xmax=2.5,ymin=-2,ymax=3,height=6cm,width=5cm,no markers,axis lines=none, xtick=\empty, ytick=\empty]
				\nextgroupplot[title={\(\)}]
				\draw[-,thick](axis cs:-1,0) -- (axis cs:0,1);
				\draw[-,thick](axis cs:0,1) -- (axis cs:1,0);
				\draw[-,thick](axis cs:-1,0) -- (axis cs:1,0);
				\draw[-,thick](axis cs:-1,0) -- (axis cs:-0.2,-0.8);
				\draw[-,thick](axis cs:-0.2,-0.8) -- (axis cs:1,0);
				\draw[fill=black](axis cs:-1,0) circle[radius=0.15];
				\draw[fill=black](axis cs:1,0) circle[radius=0.15];
				\draw[fill=black](axis cs:0,1) circle[radius=0.15];
				\draw[fill=black](axis cs:-0.2,-0.8) circle[radius=0.15];
				
				\nextgroupplot[title={\(\)}]
				\draw[-,thick](axis cs:-1,0) -- (axis cs:0,1);
				\draw[-,thick](axis cs:0,1) -- (axis cs:1,0);
				\draw[-,thick](axis cs:-1,0) -- (axis cs:1,0);
				\draw[-,thick](axis cs:-1,0) -- (axis cs:-0.2,-0.8);
				\draw[-,thick](axis cs:-0.2,-0.8) -- (axis cs:1,0);
				\draw[-,thick,blue](axis cs:-0.2,-0.8) -- (axis cs:2,-0.2);
				\draw[-,thick,blue](axis cs:0,1) -- (axis cs:2,-0.2);
				\draw[fill=black](axis cs:-1,0) circle[radius=0.15];
				\draw[fill=black](axis cs:1,0) circle[radius=0.15];
				\draw[fill=black](axis cs:0,1) circle[radius=0.15];
				\draw[fill=black](axis cs:-0.2,-0.8) circle[radius=0.15];
				\draw[fill=blue](axis cs:2,-0.2) circle[radius=0.15];
                \node[] at (axis cs:0.0, -1.3) {\LARGE$v_1$};
                \node[] at (axis cs:2.0, 0.3) {\LARGE$v_2$};

				\nextgroupplot[title={\(\)}]
				\draw[-,thick](axis cs:-1,0) -- (axis cs:0,1);
				\draw[-,thick](axis cs:0,1) -- (axis cs:1,0);
				\draw[-,thick](axis cs:-1,0) -- (axis cs:1,0);
				\draw[-,thick](axis cs:-1,0) -- (axis cs:-0.2,-0.8);
				\draw[-,thick,dashed](axis cs:-0.2,-0.8) -- (axis cs:1,0);
				\draw[-,thick,blue](axis cs:-0.2,-0.8) -- (axis cs:2,-0.2);
				\draw[-,thick,blue](axis cs:0,1) -- (axis cs:2,-0.2);
				\draw[-,thick,blue](axis cs:1,0) -- (axis cs:2,-0.2);
				\draw[fill=black](axis cs:-1,0) circle[radius=0.15];
				\draw[fill=black](axis cs:1,0) circle[radius=0.15];
				\draw[fill=black](axis cs:0,1) circle[radius=0.15];
				\draw[fill=black](axis cs:-0.2,-0.8) circle[radius=0.15];
				\draw[fill=blue](axis cs:2,-0.2) circle[radius=0.15];
                \node[] at (axis cs:0.0, -1.3) {\LARGE$v_1$};
                \node[] at (axis cs:2.0, 0.3) {\LARGE$v_2$};
			\end{groupplot}
			\tikzset{SubCaption/.style={
					text width=5cm,yshift=5mm, align=center,anchor=north}}
			\node[SubCaption] at (ch1plots c1r1.south) {\subcaption{graph $G_i$ }};
			\node[SubCaption] at (ch1plots c2r1.south) {\subcaption{Henneberg type I}};
			\node[SubCaption] at (ch1plots c3r1.south) {\subcaption{Henneberg type II}};
		\end{tikzpicture}
	\end{center}
	\caption{{\it Henneberg sequence} for (a) graph $G_i$ (b) graph $G_{i+1} $ adding vertex (blue dot) using type I (c) graph $G_{i+1} $ adding vertex (blue dot) using type II (dashed line represents edge will be removed).}
	\label{fig-Henneberg-2d}	
\end{figure}
As we can see in Fig. \ref{fig-Henneberg-2d} (a), the graph $G_i$ can be constructed from gluing two triangles. Then, in Fig. \ref{fig-Henneberg-2d} (b), $I_2$ from $v_2$ can be attached to add the new vertex. This new vertex does not have to be a triangle. For Fig. \ref{fig-Henneberg-2d} (b), there is new tetragon generated when $I_2$ is added. For type II, this is just rearrangement of the edge, Notice that this operation does not change the number of triangles and tetragons. For the three-dimensional case, the process is very similar to two-dimensionl case except for using $I_3$ instead of $I_2$ in the type I operation. More details about these sequences can be found in \cite{TayWhiteley1985}.
\begin{definition}[{\cite{Bartzos20}}\cite{Grasegger18}\cite{Borcea04}]
Henneberg sequences are used to construct the Geiringer graph inductively, such that for a graph, $G$ is a sequence $G_3, G_4,\dots, G_n$ of Geiringer graphs on $3,4,\dots,n$ vertices. $G_3$ is a triangle, $G_n=G$ and each graph $G_{i+1}$ is obtained from the previous one $G_i$ via one of two steps.
	\begin{enumerate}
		\item Type I $\colon$ adds a new vertex and three new edges connecting this vertex to three arbitrary vertices of $G_i$
		\item Type II $\colon$ adds a new vertex and four new edges and removes an old edge. The four new edges must connect the new vertex to four old vertices, so at least three of them are joined via an edge that will be removed.
		\item Type III $\colon$ adds a new vertex and five new edges and removes two old edges.
	\end{enumerate}
\end{definition}
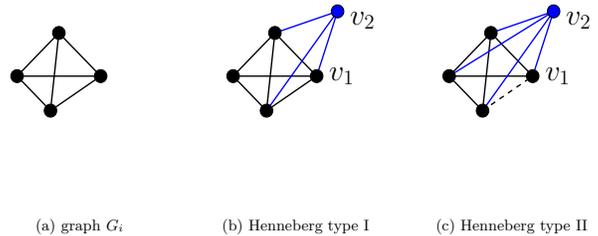
\begin{figure}[h]
	\begin{center}		
		\begin{tikzpicture}[scale=0.65, transform shape]
			\begin{groupplot}[group style={group name=ch1plots, group size=3 by 1},xmin=-1.5,xmax=2.5,ymin=-2,ymax=3,height=6cm,width=5cm,no markers,axis lines=none, xtick=\empty, ytick=\empty]
				\nextgroupplot[title={\(\)}]
				\draw[-,thick](axis cs:-1,0) -- (axis cs:0,1);
				\draw[-,thick](axis cs:0,1) -- (axis cs:1,0);
				\draw[-,thick](axis cs:-1,0) -- (axis cs:1,0);
				\draw[-,thick](axis cs:-1,0) -- (axis cs:-0.2,-0.8);
				\draw[-,thick](axis cs:-0.2,-0.8) -- (axis cs:1,0);
				\draw[-,thick](axis cs:-0.2,-0.8) -- (axis cs:0,1);
				\draw[fill=black](axis cs:-1,0) circle[radius=0.15];
				\draw[fill=black](axis cs:1,0) circle[radius=0.15];
				\draw[fill=black](axis cs:0,1) circle[radius=0.15];
				\draw[fill=black](axis cs:-0.2,-0.8) circle[radius=0.15];
				
				\nextgroupplot[title={\(\)}]
				\draw[-,thick](axis cs:-1,0) -- (axis cs:0,1);
				\draw[-,thick](axis cs:0,1) -- (axis cs:1,0);
				\draw[-,thick](axis cs:-1,0) -- (axis cs:1,0);
				\draw[-,thick](axis cs:-1,0) -- (axis cs:-0.2,-0.8);
				\draw[-,thick](axis cs:-0.2,-0.8) -- (axis cs:1,0);
				\draw[-,thick](axis cs:-0.2,-0.8) -- (axis cs:0,1);
				\draw[-,thick,blue](axis cs:-0.2,-0.8) -- (axis cs:1.5,1.5);
				\draw[-,thick,blue](axis cs:0,1) -- (axis cs:1.5,1.5);
				\draw[-,thick,blue](axis cs:1,0) -- (axis cs:1.5,1.5);
				\draw[fill=black](axis cs:-1,0) circle[radius=0.15];
				\draw[fill=black](axis cs:1,0) circle[radius=0.15];
				\draw[fill=black](axis cs:0,1) circle[radius=0.15];
				\draw[fill=black](axis cs:-0.2,-0.8) circle[radius=0.15];
				\draw[fill=blue](axis cs:1.5,1.5) circle[radius=0.15];
				\node[] at (axis cs:1.6, 0) {\LARGE$v_1$};
                \node[] at (axis cs:2.1, 1.3) {\LARGE$v_2$};

				\nextgroupplot[title={\(\)}]
				\draw[-,thick](axis cs:-1,0) -- (axis cs:0,1);
				\draw[-,thick](axis cs:0,1) -- (axis cs:1,0);
				\draw[-,thick](axis cs:-1,0) -- (axis cs:1,0);
				\draw[-,thick](axis cs:-1,0) -- (axis cs:-0.2,-0.8);
				\draw[-,thick,dashed](axis cs:-0.2,-0.8) -- (axis cs:1,0);
				\draw[-,thick](axis cs:-0.2,-0.8) -- (axis cs:0,1);
				\draw[-,thick,blue](axis cs:-0.2,-0.8) -- (axis cs:1.5,1.5);
				\draw[-,thick,blue](axis cs:0,1) -- (axis cs:1.5,1.5);
				\draw[-,thick,blue](axis cs:1,0) -- (axis cs:1.5,1.5);
				\draw[-,thick,blue](axis cs:-1,0) -- (axis cs:1.5,1.5);
				\draw[fill=black](axis cs:-1,0) circle[radius=0.15];
				\draw[fill=black](axis cs:1,0) circle[radius=0.15];
				\draw[fill=black](axis cs:0,1) circle[radius=0.15];
				\draw[fill=black](axis cs:-0.2,-0.8) circle[radius=0.15];
				\draw[fill=blue](axis cs:1.5,1.5) circle[radius=0.15];
				\node[] at (axis cs:1.6, 0) {\LARGE$v_1$};
                \node[] at (axis cs:2.1, 1.3) {\LARGE$v_2$};
			\end{groupplot}
			\tikzset{SubCaption/.style={
					text width=5cm,yshift=-8mm, align=center,anchor=north}}
			\node[SubCaption] at (ch1plots c1r1.south) {\subcaption{graph $G_i$ }};
			\node[SubCaption] at (ch1plots c2r1.south) {\subcaption{Henneberg type I}};
			\node[SubCaption] at (ch1plots c3r1.south) {\subcaption{Henneberg type II}};
		\end{tikzpicture}
	\end{center}
	\caption{{\it Henneberg sequences} (a) graph $G_i$ (b) graph $G_{i+1} $ adding vertex (blue dot) using type I (c) graph $G_{i+1} $ adding vertex (blue dot) using type II (dashed line represents edge will be removed).}
\label{fig-Henneberg-3d}
\end{figure}
However, as we observed from the example, this sequences cannot be applied to make redundantly rigid graphs. For example, Fig. \ref{fig-non-Henneberg} shows adding a new edge to a rigid graph in 2D and 3D. Since these new edges will use previous vertices, one can easily check by examining edges whether it is made from vertex members in the rigid graph. Note that we do not include Henneberg type III operations since there are some known flexible cases as in \cite{Grasegger2020}. Instead, we will define edge rearrangement sequence where the edge is relocated to different place while it is preserving the rank of rigidity matrix.
\begin{figure}
		\captionsetup{singlelinecheck = false, justification=raggedright}
		\begin{center}
  \begin{tabular}{c c}
     \includegraphics[height=0.25\textwidth]{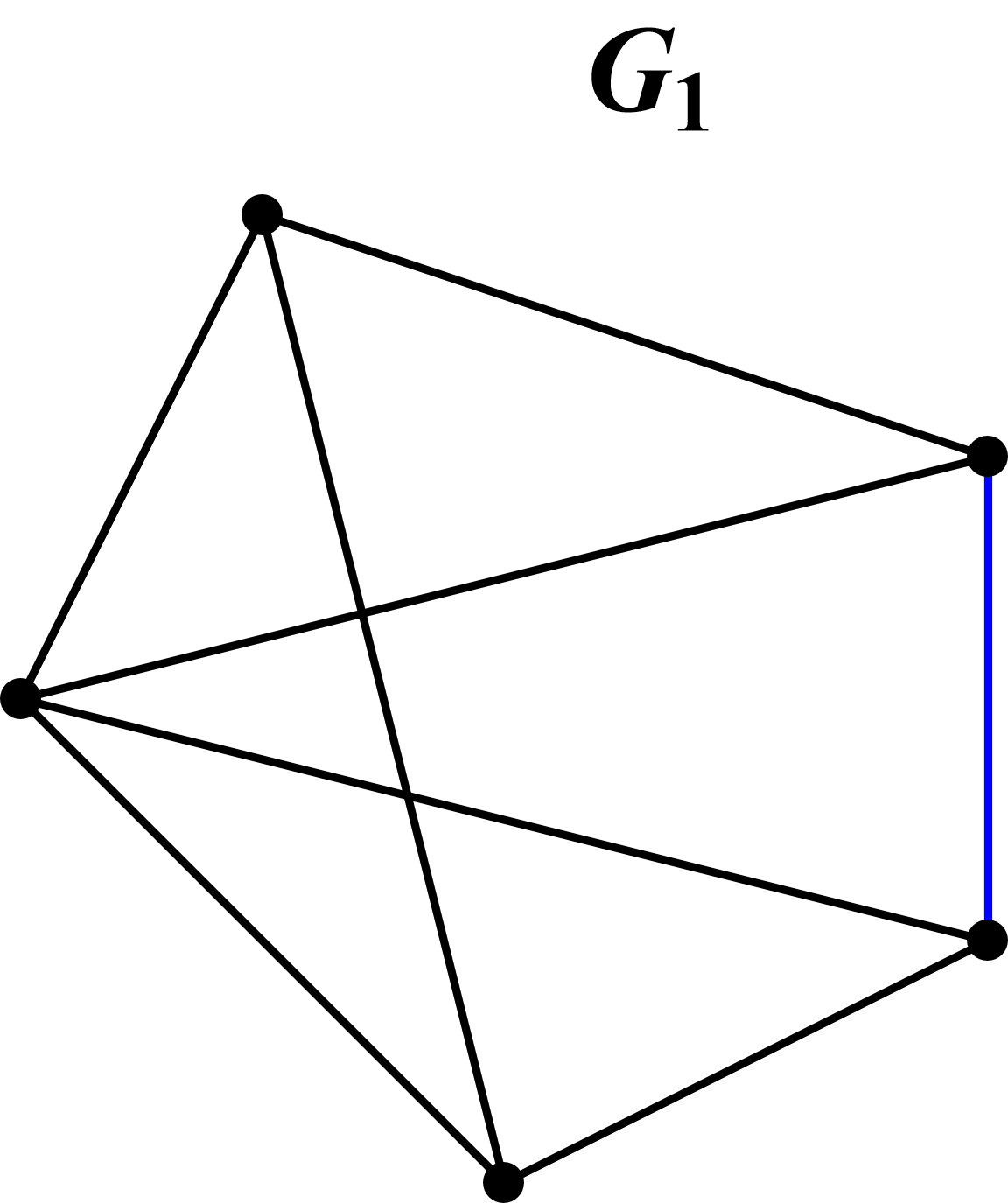}  &
     \includegraphics[height=0.28\textwidth]{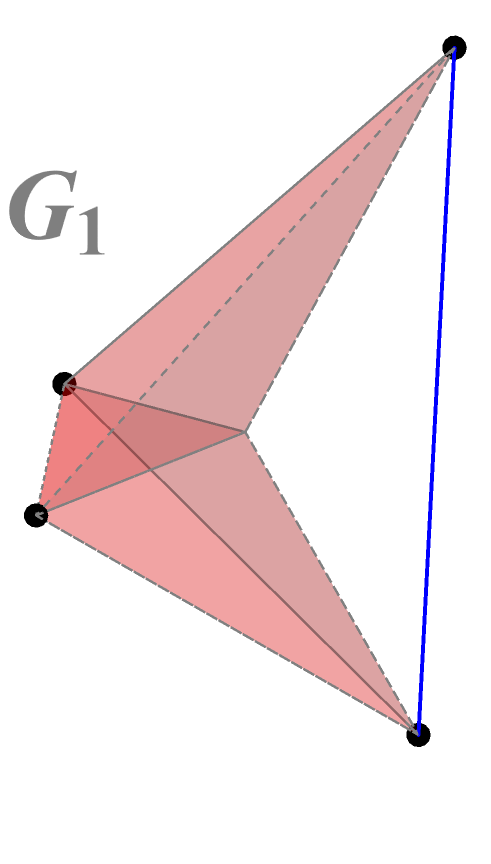}\\
       (a) & (b) 
  \end{tabular}    
	\caption{{\it Non-Henneberg type operation:} (a) Adding an edge to existing graph in 2D. (b) Adding an edge to existing graph in 3D.}
\label{fig-non-Henneberg}
 \end{center}	
\end{figure}
Therefore, to construct rigid graphs, we consider the following steps:
\begin{itemize}
    \item Simplex clusters,
    \item Adding $I_d$ to the rigid clusters,
    \item Adding extra edges on the rigid clusters,
    \item $(d-1)$-simplex loops (itself or around the rigid graph) with $(d+1)$-connected.
\end{itemize}
Notice that the first two cases are the same as Henneberg sequence type I and II. For the third item, since these extra edges can be included after vertices constructing the minimally rigid graphs identified, edges can be taken by $E(V)$ (taking all edges in between rigid vertices). Thus, for the algorithm, we can save rigid vertex information of the graph and edges can be chosen based on these vertices which will cost $\mathcal{O}(|V|)$ from the connectivity list. Next, we will talk more details for the last case. This is the combination of $C_{d-1}$ and $(d+1)$-connected. This structure comes from the combination of the type I and type II. For example, Fig. \ref{fig-Henneberg-2dex} (a) shows a rigid graph constructed by five triangles ($2$-simplex). Then, by using type II sequence, we can move four edges $(p_2,p_5),(p_3,p_6),(p_4,p_7),(p_1,p_8)$ to $(p_5,p_6),(p_6,p_7),(p_7,p_8),(p_5,p_8)$ which corresponds to Fig. \ref{fig-Henneberg-2dex} (b). The dashed part in Fig. \ref{fig-Henneberg-2dex} (b) constructs $1$-simplex loop around the rigid graph of $(p_1\sim p_4)$. We can assume that if the graph is constructed from Henneberg type I and II operations, the graph will maintain its rigidity as long as any operations with this sequence are maintaining the same rank of the rigidity matrix. 
\begin{figure}
		\captionsetup{singlelinecheck = false, justification=raggedright}
		\begin{center}
  \begin{tabular}{c c}
     \includegraphics[height=0.25\textwidth]{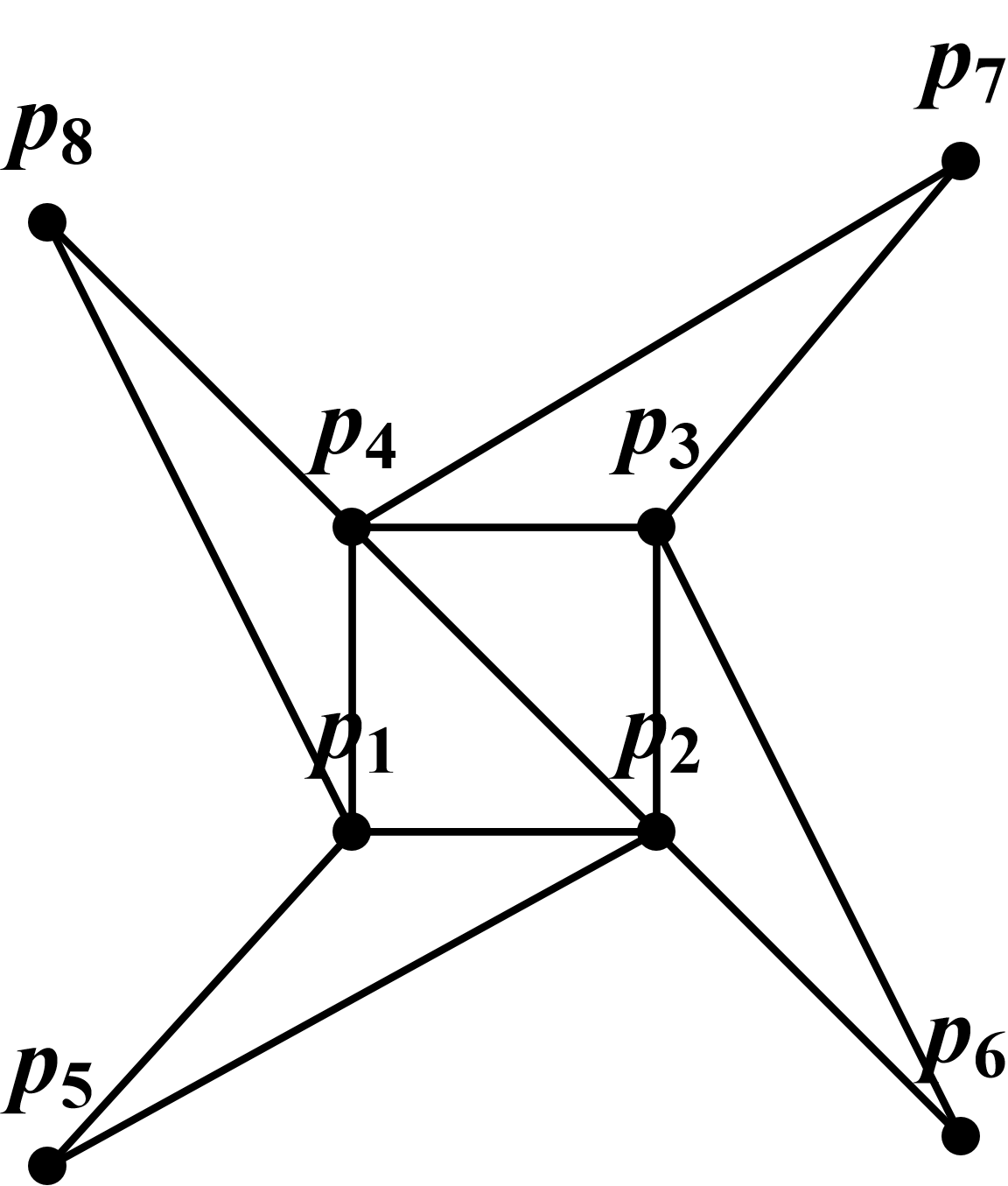}  &
     \includegraphics[height=0.25\textwidth]{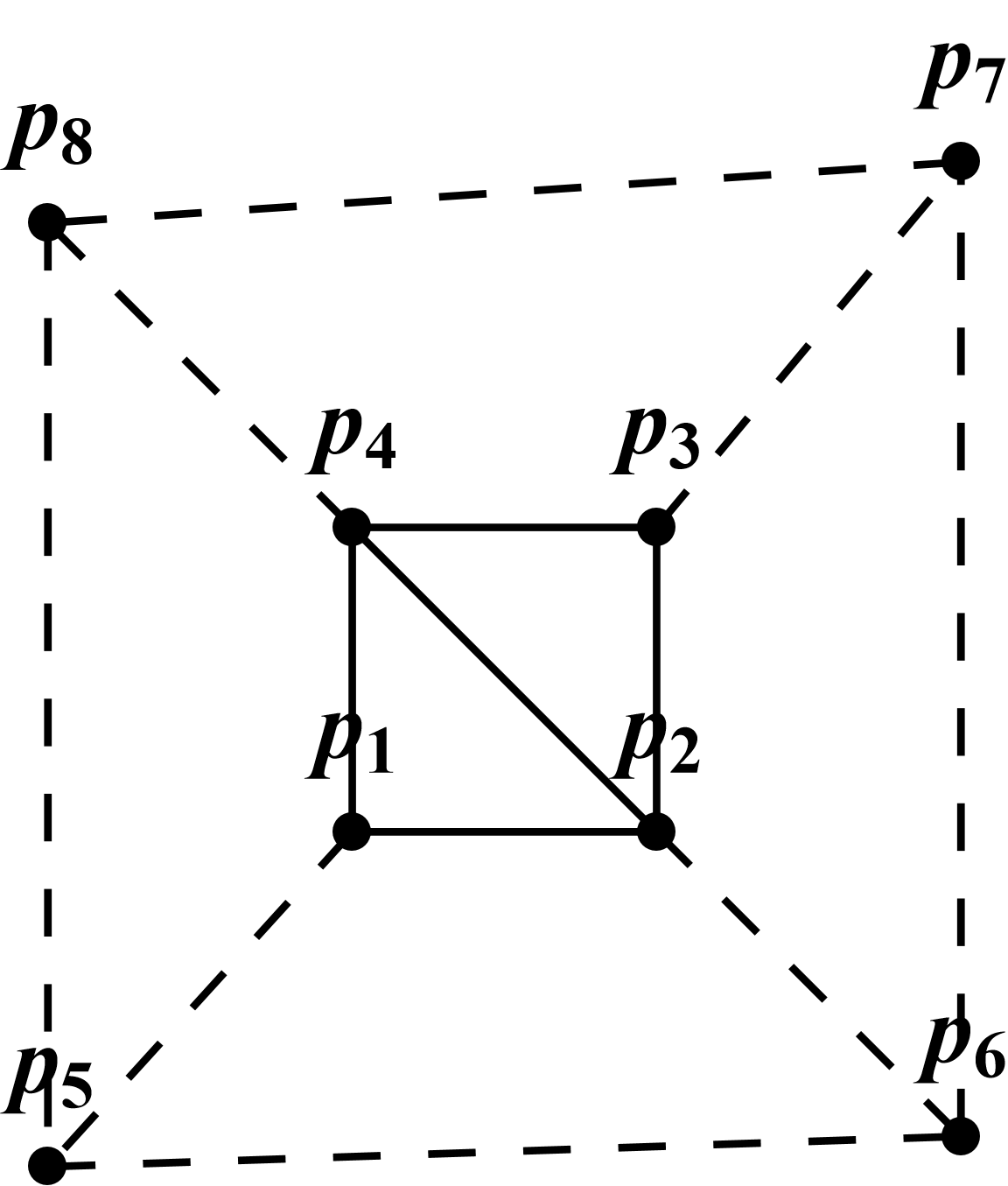}\\
       (a) & (b) 
  \end{tabular}    
	\caption{{\it Henneberg type I to II operation (2D):} (a) Adding type I vertices $(p_5\sim p_8)$ to existing graph $(p_1\sim p_4)$ in 2D. (b) Move four edges from (a) to construct $1$-simplex ($C_1$ in Definition \ref{defsimplex}) loops (dashed black lines) around rigid graph (solid black lines).}
\label{fig-Henneberg-2dex}
 \end{center}	
\end{figure}
In three dimensions, this process can create a triangulated surface, and there has been extensive research on the rigidity of such surfaces. Euler initially conjectured that closed surfaces are rigid. Later, Cauchy (1813) showed that a closed, strictly convex polyhedron must be rigid \cite{Cauchy1813}, and Gluck (1975) demonstrated that almost all simply connected, closed surfaces are rigid \cite{Gluck1975}. However, Connelly (1979) \cite{Connelly1979} showed that not all triangulated surfaces are rigid. Thus, without looking a rigidity matrix, it may be difficult to see whether the triangulated surface is rigid or not. However, if we recall Henneberg sequences, there is a trick to check. Since from Henneberg type I to II changed simplex cluster to one with $I_d$ or $I_{d-1}$ structures, by applying from II to I, we can see the one of possible graph structures with maximal number of simplices. Although there could be multiple structures possible by this process, the rigidity of the graph would not be changed as long as they are generic sets. Therefore, we will define maximal simplical graph in here.
\begin{definition}
    For graph $G$, edge rearrangement sequence is the operation of relocating the edge $e_i$ connected to the pair of vertices $(v_i,v_j)$ to $(v_k,v_l)$ for $v_i\neq v_k,v_j\neq v_k,v_l$ to construct the new graph $G'$ while maintaining the rank of the rigidity matrix such that $\rank \mathcal{R}(G,p)=\rank \mathcal{R}(G',p)$.
\end{definition}
\begin{definition}
    We define the function $F_{C_d}$ to get maximal simplicial graph $G'$ of graph $G$ via $G'=F_{C_d}(G)$ which is the graph structure can be derived from $G$ by applying from Henneberg II to I (or by edge rearrangement sequence) such that $G'$ have the largest number of $C_d$ simplices in $d-$dimensional space without changing the number of edges and vertices, and also maintaining the rank of rigidity matrix.
\end{definition}
For estimating the number of $C_d$ simplices in the graph $G$, we approximate the number of $d$-dimensional faces by $\text{number of faces}/2$. This approximation comes from considering the $3$ faces required to make $C_2$ or $C_3$ simplices. After accounting for a shared face, we need only two additional faces to construct $C_2$ or $C_3$ simplices, excluding internal faces for triangulated surfaces in 3D. For example, in Fig. \ref{fig-Henneberg-2dex} (b), there are eight edges that do not contribute to forming triangles. Since edges are faces in the triangle as a $C_2$ simplex, we estimate that approximately $4$ triangles can be derived by rearranging these edges. Similarly, Fig. \ref{fig-Henneberg-3dex} (b) shows $12$ faces and no tetrahedron. The rank of the rigidity matrix for $G_1$ is $18$, as calculated from the coordinate vectors in Fig. \ref{fig-Henneberg-3dex} (b), meaning this graph is rigid since $3\times 8-6=18$. With $12$ faces, we estimate that $G_1$ can be rearranged into $6$ tetrahedrons by simple counting, but it turns out to be $5$ as shown in Fig. \ref{fig-Henneberg-3dex} (c), due to shared bottom faces ($10/2=5$).

In Fig. \ref{fig-Henneberg-3dex} (c), all tetrahedrons share three vertices, confirming that they don’t require additional edges, according to the rules outlined in the previous section. Moving to the next example, consider the octahedron $G_2$ as shown in Fig. \ref{fig-Henneberg-3dex} (d). The rigidity matrix $\mathcal{R}(G_2,p)$ for the octahedron is:
\[
\begin{bmatrix}
    p_1-p_2&p_2-p_1&0&0&0&0\\
    p_1-p_3&0&p_3-p_1&0&0&0\\
    p_1-p_4&0&0&p_4-p_1&0&0\\
    p_1-p_5&0&0&0&p_5-p_1&0\\
    0&p_2-p_3&p_3-p_2&0&0&0\\
    0&0&p_3-p_4&p_4-p_3&0&0\\
    0&0&0&p_4-p_5&p_5-p_4&0\\
    0&p_2-p_5&0&0&p_5-p_2&0\\
    0&p_2-p_6&0&0&0&p_6-p_2\\
    0&0&p_3-p_6&0&0&p_6-p_3\\
    0&0&0&p_4-p_6&0&p_6-p_4\\
    0&0&0&0&p_5-p_6&p_6-p_5
\end{bmatrix}
\].
Here, we use the simpler notation $p_1-p_2$ to represent $(p_1-p_2)_x,(p_1-p_2)_y,(p_1-p_2)_z$. The rank of this matrix is $12$. As we move from Fig. \ref{fig-Henneberg-3dex} (c) to (d), the row vector $(0, p_2-p_5, 0, 0, p_5-p_2, 0)$ changes to $(p_1-p_6, 0, 0, 0, 0, p_6-p_1)$. If this new row vector is independent of the other rows, then the matrix rank remains unchanged. By changing the edge $(p_2,p_5)$ to $(p_1,p_6)$, two tetrahedrons join via two vertices, and we observe that edge $(p_3,p_4)$ connects the two tetrahedrons, ensuring that the graph is rigid. We verify this by comparing the rank of the rigidity matrix: $3\times 6-6=12$. Since the graph has $8$ faces, we expect $4$ tetrahedrons. However, internal faces reduce this to $3$, as shown in Fig. \ref{fig-Henneberg-3dex} (f).

This observation implies that we can check the rigidity of the graph without rearranging the edges to satisfy the maximal number of simplices. Since we are interested in verifying whether the original graph is rigid, we can alter the position of edges to confirm rigidity. Changing edge positions may result in different graph realizations based on the given edge lengths.
\begin{figure}
		\captionsetup{singlelinecheck = false, justification=raggedright}
		\begin{center}
  \begin{tabular}{c c c}
     \includegraphics[height=0.12\textwidth]{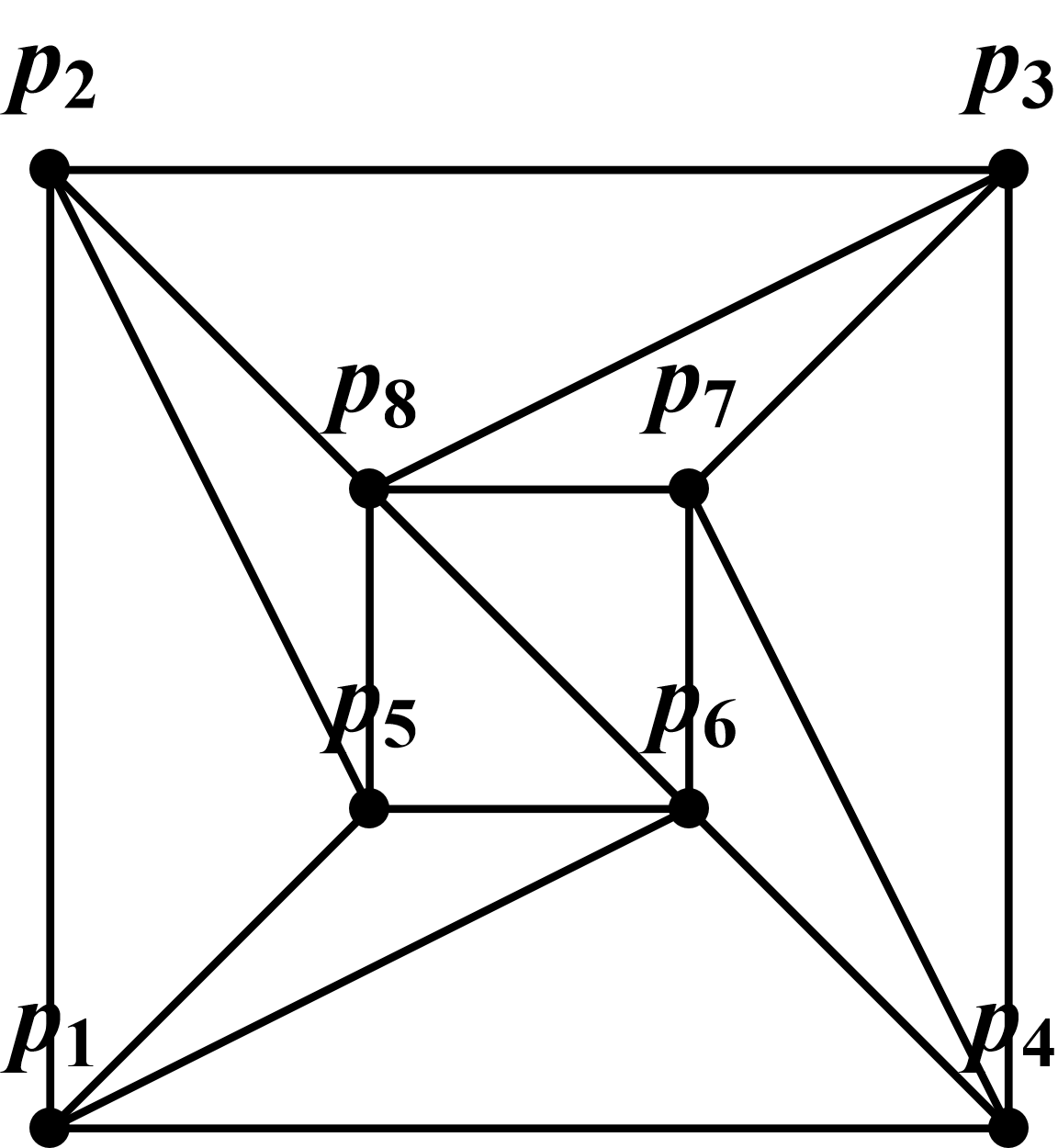} &
     \includegraphics[height=0.14\textwidth]{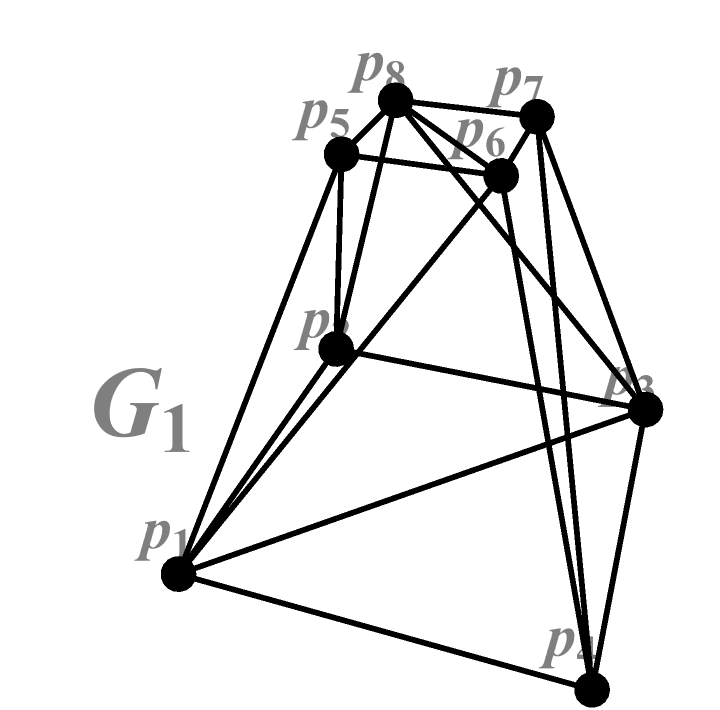} &
     \includegraphics[height=0.14\textwidth]{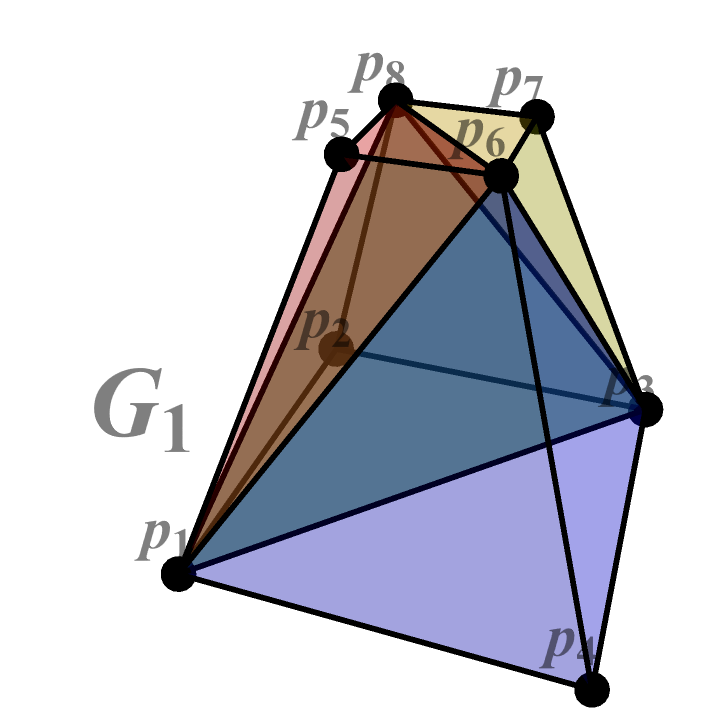}\\
       (a) & (b) & (c)\\
     \includegraphics[height=0.14\textwidth]{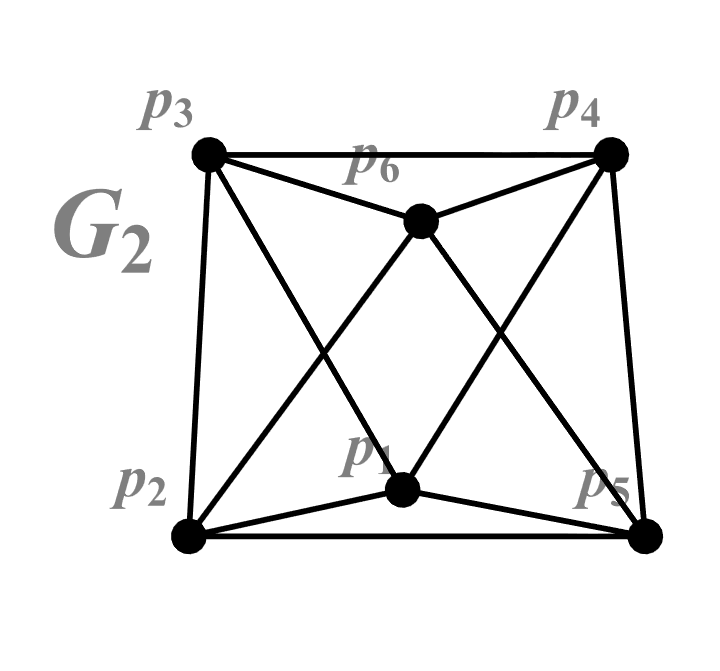} &
     \includegraphics[height=0.14\textwidth]{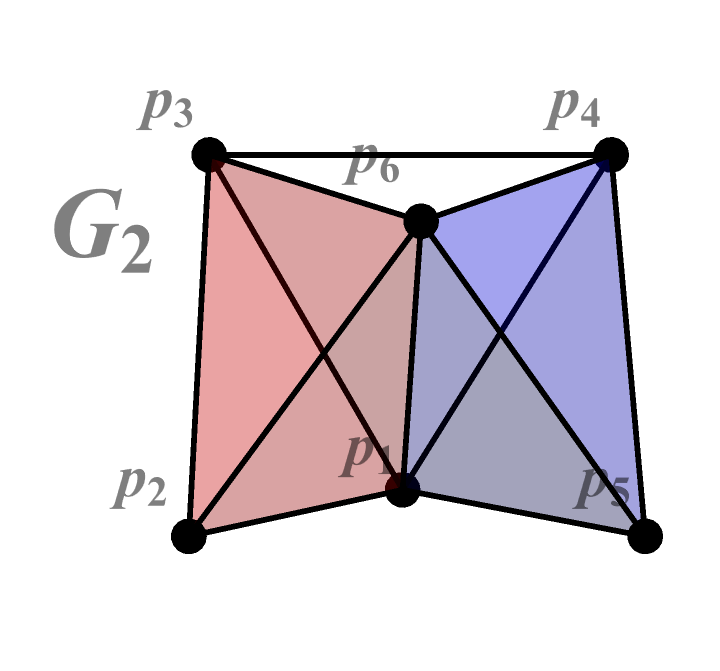} &
     \includegraphics[height=0.14\textwidth]{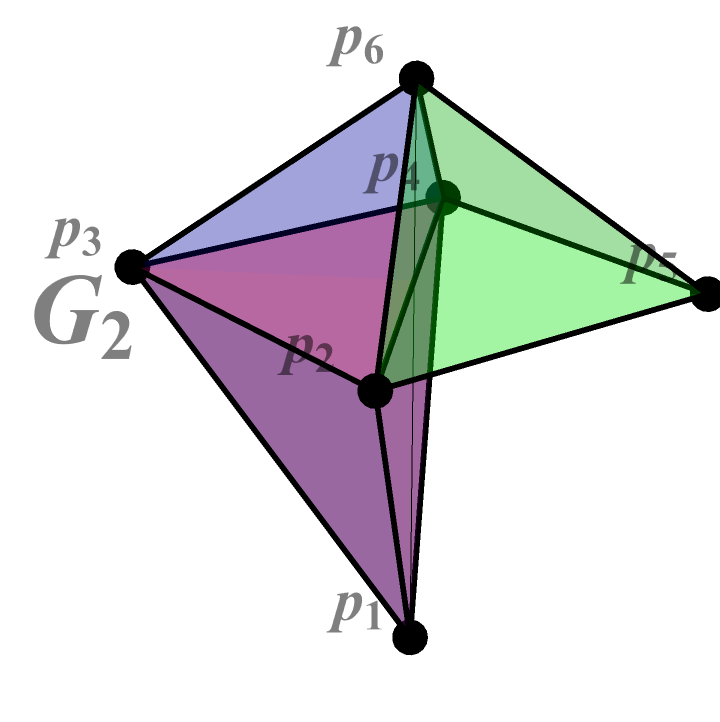} \\
       (d) & (e) & (f)
  \end{tabular}    
	\caption{{\it Henneberg Type II to I operation (3D):} (a) Top view of $G_1$ (edge $(p_1,p_3)$ not visible). (b) Triangulated surface of $G_1$. (c) Graph $F_{C_2}(G_1)$ after rearranging edges $(p_4,p_7)$ and $(p_2,p_5)$ in (b). (d) Octahedron $G_2$. (e) Graph after rearranging edge $(p_2,p_5)$ to $(p_1,p_6)$ in (d). (f) $F_{C_2}(G_2)$ for $G_2$.} 
\label{fig-Henneberg-3dex}
 \end{center}	
\end{figure}
In performing the Henneberg Type II to I operation in 3D, it is easiest to check whether the triangles are oriented similarly. For example, in Fig. \ref{fig-3dex} (a), three triangles point upwards $(p_6)$ and three point downwards $(p_5)$. Although these six faces could form at least one tetrahedron, the orientation prevents this. By rearranging edges, we can construct a graph with tetrahedrons. Moving the edge $(p_2,p_3)$ to the position of $(p_5,p_6)$ yields a graph with two tetrahedrons, as shown in Fig. \ref{fig-3dex} (b). The rigidity matrix $\mathcal{R}(G,p)$ for this graph is:
\[
\begin{bmatrix}
    p_1-p_2&p_2-p_1&0&0&0&0\\
    p_1-p_5&0&0&0&p_5-p_1&0\\
    p_1-p_6&0&0&0&0&p_6-p_1\\
    0&p_2-p_3&p_3-p_2&0&0&0\\
    0&p_2-p_5&0&0&p_5-p_2&0\\
    0&p_2-p_6&0&0&0&p_6-p_2\\
    0&0&p_3-p_4&p_4-p_3&0&0\\
    0&0&p_3-p_5&0&p_5-p_3&0\\
    0&0&p_3-p_6&0&0&p_6-p_3\\
    0&0&0&p_4-p_5&p_5-p_4&0\\
    0&0&0&p_4-p_6&0&p_6-p_4
\end{bmatrix}
\]
When both the row vector $(0,p_2-p_3,p_3-p_2,0,0,0)$ and the row vector $(0,0,0,0,p_5-p_6,p_6-p_5)$ are linearly independent compared to other rows in the rigidity matrix $\mathcal{R}(G,p)$, replacing the first row with the second does not affect the matrix's rank. This suggests that, for any graph with $|E| \geq S(n,d)$ edges, if the edges are positioned such that their corresponding row vectors are linearly independent in the rigidity matrix for any of the $S(n,d)$ combinations, swapping edges does not alter the matrix's rank.

This insight allows us to develop an algorithm that systematically checks the rigidity of a graph by counting the number of triangles and adjusting the orientation of edges. Specifically, when more than three connected triangles fail to form a tetrahedron, we flip the orientation of the edges to ensure rigidity.

In 2D, a similar operation can be applied by moving $I_3$ subgraph loops such that at least two of their members are connected within the rigid cluster. This ensures that edge rearrangements maintain the rank of the rigidity matrix, thereby preserving the rigidity of the overall structure.
\begin{figure}
		\captionsetup{singlelinecheck = false, justification=raggedright}
		\begin{center}
  \begin{tabular}{c c}
     \includegraphics[height=0.2\textwidth]{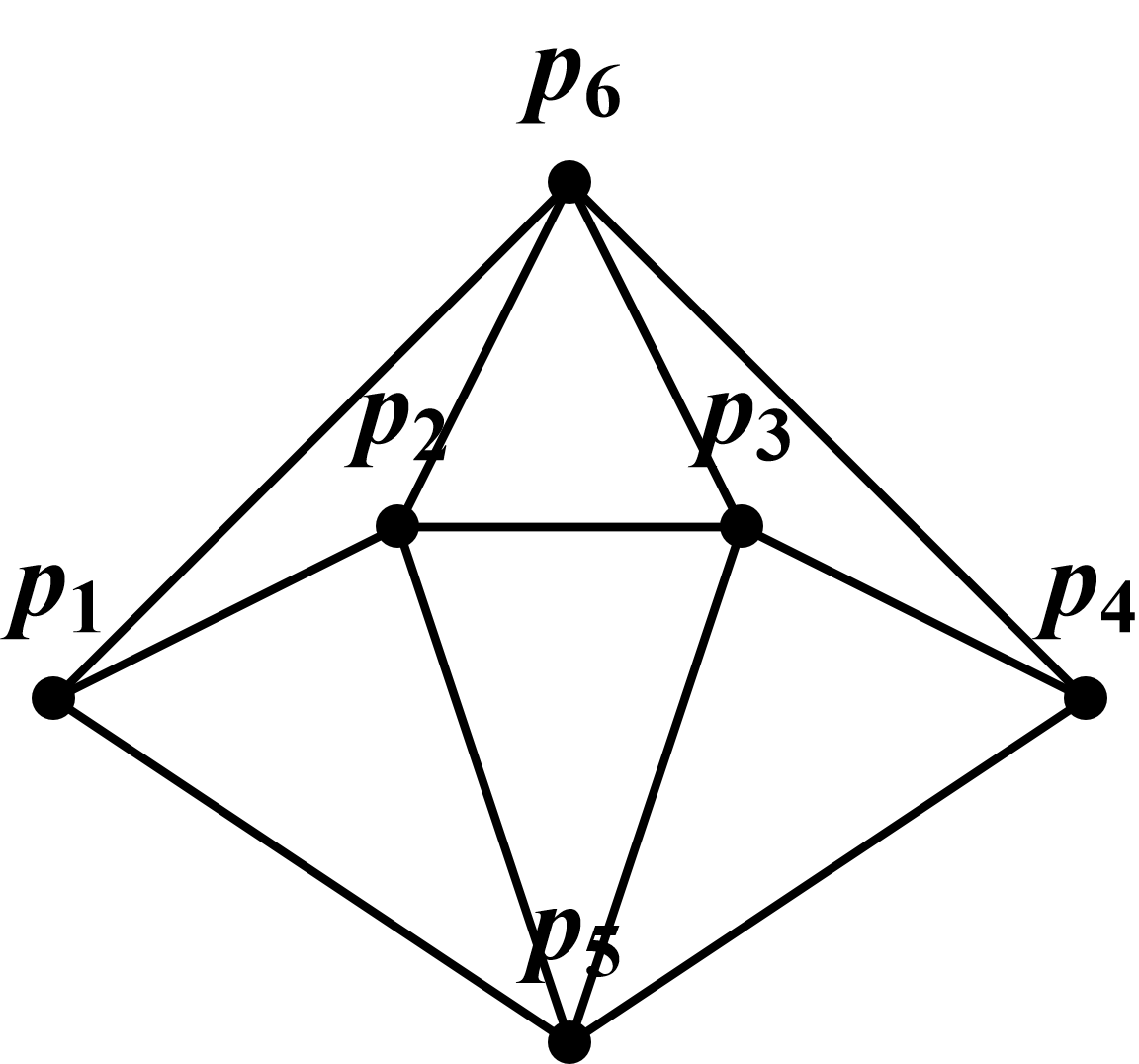}  &
     \includegraphics[height=0.2\textwidth]{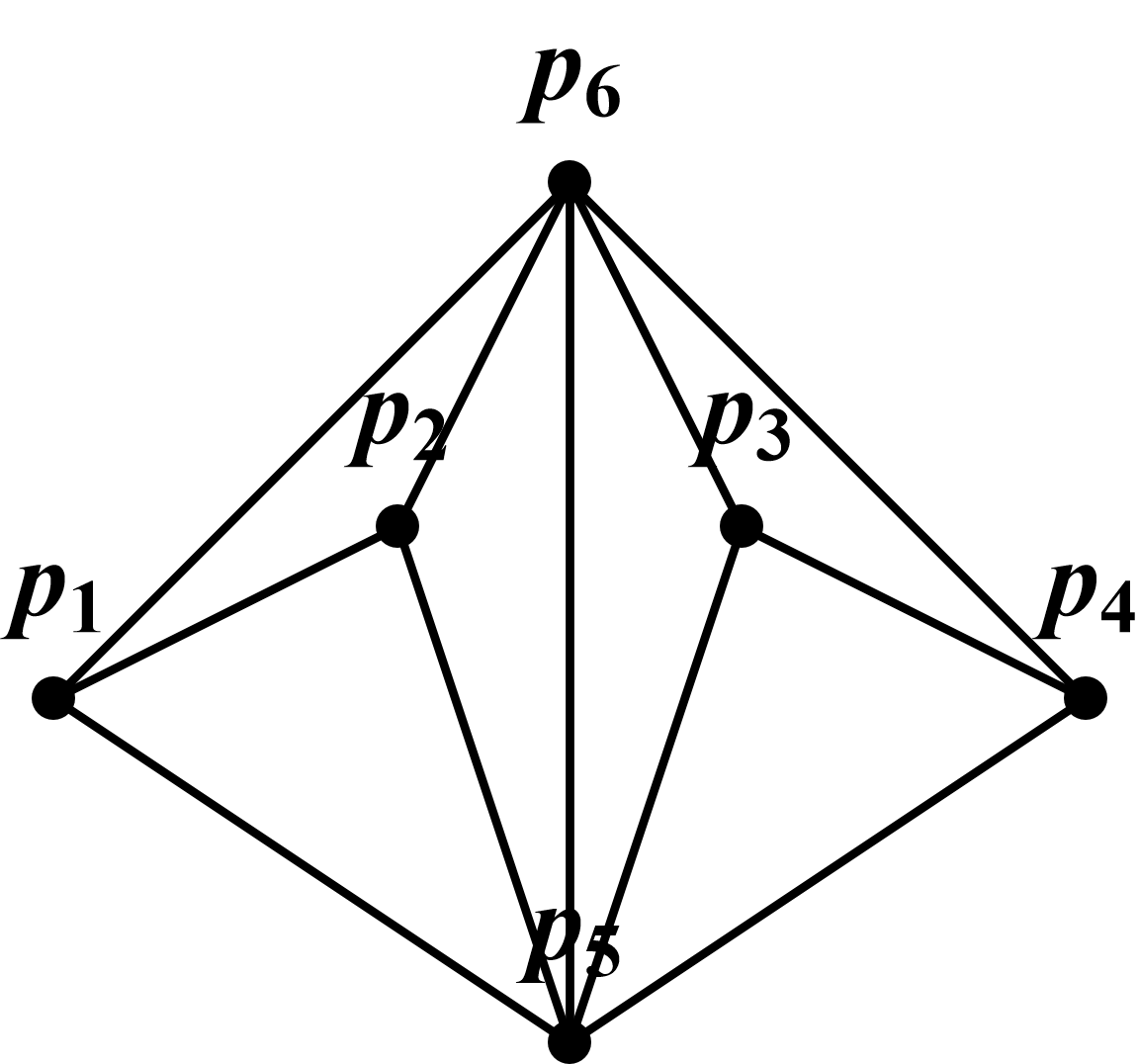}\\
       (a) & (b) 
  \end{tabular}    
	\caption{{\it Edge rearrangement sequence:} (a) The graph with six faces in $3$d. (b) The edge $(p_2,p_3)$ can be rotated and located to $(p_5,p_6)$.}
\label{fig-3dex}
 \end{center}	
\end{figure}
\subsection{Algorithms to check rigidity by its connection in $\mathbb{R}^d$}
In this section, we discuss methods for checking the rigidity of the generic graph. To achieve this, we decompose graphs into their smallest rigid units, similar to Fig. 2 in \cite{Berg2003}, but applicable to both minimally and redundantly rigid graphs. We then check the connectivity between these units, gradually building up larger parts to identify rigid clusters. Moukarzel used body-bar rigidity to assess graph rigidity in 2D \cite{Moukarzel1996} by matching subsets of edges. Our approach shares some similarities by treating a rigid graph as a body; however, we introduce an explicit counting between two rigid graphs without checking the paths.

Since we assume generic configurations, we do not account for algebraically dependent vertex sets. This ensures that the graph $G$ achieves the maximal rank possible in the rigidity matrix, indicating whether the connections are rigid, but not necessarily implying rigidity for any specific realization of the graph. Moreover, there are cases where generic configurations cannot be assumed, such as in experimental settings. In these instances, an algorithm can be inserted to check the rigidity matrix (or pinned rigidity matrix) for each rigid cluster or simplex.

While we explain a procedure for assessing rigidity without using the rigidity matrix, depending on the size and structure of the graph, one can always use the rigidity matrix to verify subgraphs and expedite the process. This algorithm is based on the results of the previous section, assuming all $C_d$ components are rigid by themselves. We begin by identifying $C_d$s within the graph $G$. As previously mentioned, $C_{d-1}$ loops that are $(d+1)$-connected may not form simplices, even though they might be rigid. Therefore, we convert these into structures containing $C_d$s to evaluate the rigidity of the graph.

To find triangles within the graph, several methods can be used. For instance, one can search for triangles from each vertex (``node-iterator") or each edge (``edge-iterator"), as demonstrated in \cite{Schank2005}. While we focus on the node-iterator method, other alternative algorithms, such as those in \cite{magniez2005,Schank2005,LATAPY2008,Gall2014,dumitrescu2024}, can be applied based on the graph's size and the density of its rigid clusters. 

The ``node-iterator" method tests, for each pair of neighbors, whether they are connected by an edge. For a given vertex $v$, this requires checking $\binom{d(v)}{2}$ pairs, where $d(v)$ denotes the degree, or the number of adjacent nodes, of vertex $v$, and $2$ is for a triangle (corresponding to $C_2$). The complexity of this approach can be estimated as $\mathcal{O}(n d_{\max})$, where $d_{\max}$ is the maximal degree in the graph.

Algorithm \ref{algo1} outlines the process for finding triangles (denoted as $C_2$) using the node-iterator method. This method operates on a connectivity list, $\conn(V)$, derived from the vertex set. To reduce the time required for these checks, we utilize a unidirectional connectivity list, where connections are only stored for the vertex with the smaller ID. This eliminates redundant checks across both directions, thus reducing the number of tests to approximately $\frac{1}{2} \sum \binom{d(v)}{2}$ pairs. Depending on code efficiency, both bi-directional and unidirectional connectivity lists can be maintained and applied in different contexts.

Next, the algorithm checks whether the identified pairs are connected via an edge by assigning a unique ID to each edge. One method for assigning edge IDs is based on vertex information. For example, the ID of an edge $E$ connecting vertices $v_1$ and $v_3$ can be represented either as a fixed digit combination $00010003$ (ascending order) or as the product of prime numbers $P(1) \times P(3)$, where $P(x)$ denotes the $x$-th prime number. The prime number method is effective for small graphs but may not scale well for larger graphs due to the rapid growth of prime numbers.

To illustrate, if we are searching for a pair $(v_3, v_4)$ from vertex $v_1$, we can check whether there is an edge with an ID corresponding to $P(3) \times P(4) = 35$. This method can be extended to tetrahedrons, where we need to create a list and match the edge IDs. For instance, if vertex $v_1$ is connected to the set of vertices $(v_3, v_5, v_6)$, the corresponding edge IDs for the tetrahedron are:
\begin{align*}
&\{\Fid(E(v_3, v_5)), \Fid(E(v_3, v_6)), \Fid(E(v_5, v_6))\}\\
= &\{P(3) \times P(5), P(3) \times P(6), P(5) \times P(6)\} = \{55, 65, 143\}
\end{align*}
where $\Fid(e_i)$ is the function assigning an ID to edge $e_i$ based on predefined rules. We can then check whether $\{55, 65, 143\} \cap \Fid(E) = \{55, 65, 143\}$, where $\Fid(E)$ represents the list of all edge IDs. Depending on the programming language or software used, this set-matching operation could be more efficient than looping over edge IDs, especially if there are relevant packages available to match multiple sets simultaneously. Alternatively, an edge connectivity list can also be utilized. The details of the looping method are explained further in Algorithm \ref{algo1}.

While looping over vertices to find tetrahedra, we can simultaneously search for triangles, which is unnecessary in the two-dimensional case if an edge list is already available. Additionally, as previously mentioned, we need to transform relevant $C_{d-1}$ structures into $C_d$ structures to assess rigidity without using the rigidity matrix. Although this transformation can be performed separately, it is more efficient to integrate it while looping over vertices to identify $C_d$ structures, as transformable $C_{d-1}$ structures share at least one vertex. For instance, in three-dimensional space, an edge can be relocated when it is part of two or more neighboring triangles without forming a tetrahedron. Therefore, this operation can be incorporated into the process of listing $C_d$ structures, as shown in Fig. \ref{fig:process-chart} (where edges are modified when four triangles connect via $v_i$). However, care must be taken to avoid flipping the edges of existing simplices or double-flipping.

Another method for converting $C_{d-1}$ structures to $C_d$ in three dimensions is to search for pyramid shapes rather than individual $C_{d-1}$ structures and then convert them into tetrahedra, as illustrated in Fig. \ref{fig-3dex}. This can be achieved by examining $\binom{d(v)}{4}$ neighbors instead of $\binom{d(v)}{2}$ when looping over each vertex to identify triangles. Since these pyramids are constructed from neighboring triangles, we need to search for $\frac{1}{2} \sum \binom{d(v)}{4}$, with the factor of $\frac{1}{2}$ adjustable depending on the openness and closeness of the graph. For example, Fig. \ref{fig-Henneberg-3dex}(d) shows six pyramids based on each vertex, and Fig. \ref{fig-Henneberg-3dex}(f) demonstrates how these six pyramids can be converted into three tetrahedra, while rigidity can still be checked by examining connectivity in Fig. \ref{fig-Henneberg-3dex}(e).

As mentioned earlier, when rearranging edges to form $C_d$ structures, the rigidity matrix (or related processes, such as the force matrix discussed in \cite{PELLEGRINO1986}) can be used to verify whether these edges are independent, thus removing algebraically dependent edge sets. Alternatively, instead of converting $C_{d-1}$ structures, vertices from previously identified $C_{d-1}$ or $C_d$ structures can be used as pinned vertices and checked using a pinned rigidity matrix, as defined in Lemma \ref{gluelemma3}.
After identifying all $C_d$ structures and rigid graphs, we can merge these rigid graphs into clusters, provided they are connected by enough edges $|\binom{d+1-m}{2}|$. This merging can be accomplished using a union-find algorithm to combine subsets into the maximal size, a well-known technique described in \cite{wiki:uf}. Lee {\it et al.} employed a similar function called union pair-find for rigid cluster data structures in \cite{Lee2005}, with a complexity of $\mathcal{O}(n^2)$ for $|V|=n$. The union-find function itself operates in $\mathcal{O}(m)$ for $m$ subsets (here, the number of rigid graphs), so we expect the running time to be of the order of $m$. In Fig. \ref{fig:alg-flow}, merging simplices and clusters are shown as separate processes, but the underlying algorithms are the same, as detailed in Algorithm \ref{algo2}.

Additionally, any leftover vertices not included in rigid clusters can be checked to determine whether they connect to a single cluster in the form of $I_d$. Whether we need the entire structure (i.e., the rigid graph with both rigid vertices and edges) or just the rigid vertices will dictate whether an extra step is necessary. If the entire structure is required, we can also verify whether the edges $E$ of the graph $G$ belong to the rigid vertices $V$ in $G$ to include extra edges (over-constrained), as shown in Fig. \ref{fig-non-Henneberg} (b).

For example, consider $\cup_{v_i} \conn(v_i) \cap V$ for each vertex $v_i \in V$ in $G$, and either convert this connectivity list into an edge list or compare the connectivity list of vertices in the rigid clusters of $G$ with the corresponding edge list of $G$. The entire process is summarized in Fig. \ref{fig:alg-flow}. Fig. \ref{fig:alg-sample} shows an example run of the algorithm in 2D and 3D.

\pgfdeclarepattern{
  name=hatch,
  parameters={\hatchsize,\hatchangle,\hatchlinewidth},
  bottom left={\pgfpoint{-.1pt}{-.1pt}},
  top right={\pgfpoint{\hatchsize+.1pt}{\hatchsize+.1pt}},
  tile size={\pgfpoint{\hatchsize}{\hatchsize}},
  tile transformation={\pgftransformrotate{\hatchangle}},
  code={
    \pgfsetlinewidth{\hatchlinewidth}
    \pgfpathmoveto{\pgfpoint{-.1pt}{-.1pt}}
    \pgfpathlineto{\pgfpoint{\hatchsize+.1pt}{\hatchsize+.1pt}}
    \pgfpathmoveto{\pgfpoint{-.1pt}{\hatchsize+.1pt}}
    \pgfpathlineto{\pgfpoint{\hatchsize+.1pt}{-.1pt}}
    \pgfusepath{stroke}
  }
}

\tikzset{
  hatch size/.store in=\hatchsize,
  hatch angle/.store in=\hatchangle,
  hatch line width/.store in=\hatchlinewidth,
  hatch size=5pt,
  hatch angle=0pt,
  hatch line width=.5pt,
}

\tikzstyle{startstop} = [rectangle, rounded corners, 
minimum width=3cm, 
minimum height=1cm,
text centered, 
draw=black, fill=black,text=white]

\tikzstyle{io} = [trapezium, 
trapezium stretches=true, 
trapezium left angle=70, 
trapezium right angle=110, 
minimum width=3cm, 
minimum height=1cm, text centered, 
draw=black, fill=white]

\tikzstyle{process} = [rectangle, 
minimum width=3cm, 
minimum height=1cm, 
text centered, 
text width=3cm, 
draw=black,
fill=gray,text=white]

\tikzstyle{decision} = [diamond, 
minimum width=3cm, 
minimum height=1cm, 
text centered, 
draw=black, 
fill=gray!30]
\tikzstyle{arrow} = [thick,->,>=stealth]

\begin{figure}
    \centering
\begin{tikzpicture}[node distance=2cm]

\node (start) [startstop] {Start};
\node (in1) [io, below of=start] {Graph $G$};
\node (pro1) [process, below of=in1] {Identify $C_d$s and $C_{d-1}$s of $G$};
\node (dec1) [decision, below of=pro1, yshift=-0.5cm] {Check $C_{d-1}$s};

\node (pro2a) [process, below of=dec1, yshift=-0.5cm] {Rearrange $C_{d-1}$s to form $C_{d}$s};

\node (pro2b) [process, right of=dec1, xshift=2cm] {Remove unnecessary $C_{d-1}$s from the list};
\node (out1) [io, below of=pro2a] {$C_d$s list of the graph $G$};
\node (stop) [startstop, below of=out1] {Stop};

\draw [arrow] (start) -- (in1);
\draw [arrow] (in1) -- (pro1);
\draw [arrow] (pro1) -- (dec1);
\draw [arrow] (dec1) -- node[anchor=east] {yes} (pro2a);
\draw [arrow] (dec1) -- node[anchor=south] {no} (pro2b);
\draw [arrow] (pro2b) |- (pro1);
\draw [arrow] (pro2a) -- (out1);
\draw [arrow] (out1) -- (stop);

\end{tikzpicture}
    \caption{{\it Process to obtain $F_{C_d}(G)$, or enough quantity of ${C_d}$s, to check rigidity by connections.}}
    \label{fig:process-chart}
\end{figure}
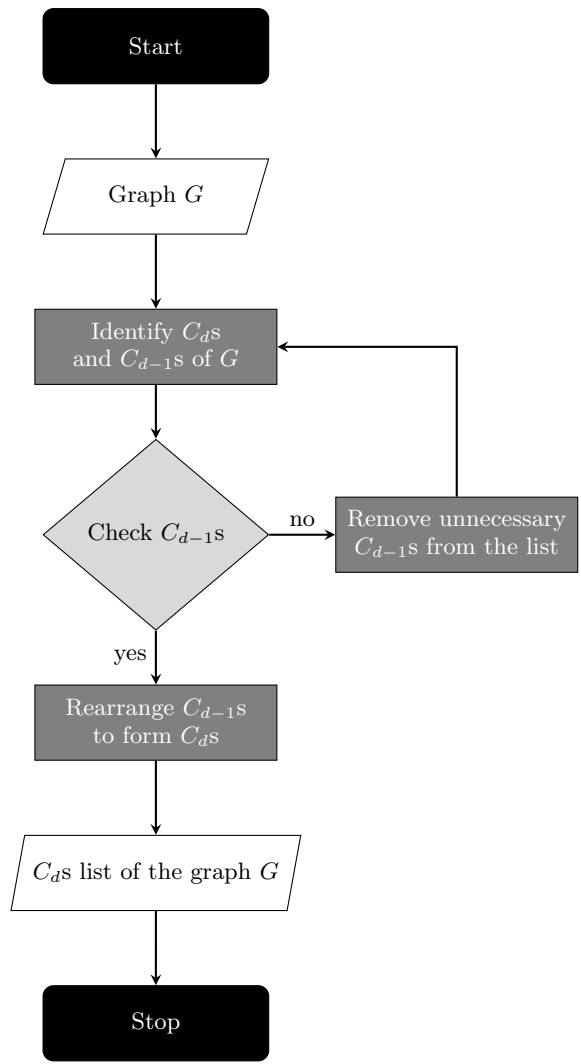

\begin{algorithm}
\begin{algorithmic}[1]
   \REQUIRE $G(V,E),\conn(V),\Fid(E)$
   \FOR[$v_i\in V$]{$v_i \leftarrow 0$ \TO $n-1$ }
       \STATE choose $d$ number of $v_j$ in the $\conn(V)$
       \STATE save $v_j$s set in the list $vlist1$
       \STATE choose $d-1$ number of $v_k$ in the $\conn(V)$ ($d>2$)
       \STATE save $v_k$s set in the list $vlist2$
       \FOR[$l\in\max(|vlist1|,|vlist2|)$]{$l \leftarrow 0$ \TO $n-1$ } 
            \IF{$\Fid(E(v_i,v_j))\in\Fid(E)$, $\forall v_j\in vlist1[l]$}
            \STATE $temp1=vlist1[l]$\COMMENT{$l$-th set is $C_d$}
            \ENDIF
            \IF{$\Fid(E(v_i,v_k))\in\Fid(E)$, $\forall v_k\in vlist2[l]$ }
            \STATE $temp2=vlist2[l]$\COMMENT{$l$-th set is $C_{d-1}$}
            \ENDIF
        \ENDFOR
         \STATE $temp2\leftarrow temp2-temp2\cap temp1$\COMMENT{Remove $C_{d-1}$s contained in $C_d$}
        \STATE $result1 \leftarrow result1 + temp1$ \COMMENT{List Addition for $C_d$}
         \STATE $result2 \leftarrow result2 + temp2$\COMMENT{List Addition for $C_{d-1}$}
   \ENDFOR
   \RETURN $result1$,$result2$
\end{algorithmic}
\caption{Algorithm to identify $C_d$ and $C_{d-1}$}
\label{algo1}
\end{algorithm}

\begin{algorithm}
\begin{algorithmic}[2]
   \REQUIRE $C_d$ list
   \WHILE[union-find]{$G_i\neq G_j$ for $G_i,G_j\in C_d$ list}
       \IF{$|V(G_i)\cap V(G_j)|=m\&|E(G_i\cup G_j)-E(G_i)-E(G_j)|=[\binom{d+1-m}{2}]$}
       \STATE list $\leftarrow G_i\cup G_j$
    \ENDIF
    \ENDWHILE
   \RETURN list
\end{algorithmic}
\caption{Algorithm to check connected clusters}
\label{algo2}
\end{algorithm}

\tikzset{
  basic box/.style={
    shape=rectangle, rounded corners, align=center, draw=#1, fill=#1!25},
  header node/.style={
    node family/width=header nodes,
    font=\strut\large,
    text depth=+.3ex, fill=white, draw},
  header/.style={
    inner ysep=+1.5em,
    append after command={
      \pgfextra{\let\TikZlastnode\tikzlastnode}
      node [header node] (header-\TikZlastnode) at (\TikZlastnode.north) {#1}
      node [span=(\TikZlastnode)(header-\TikZlastnode)]
           at (fit bounding box) (h-\TikZlastnode) {}
    }
  },
  fat black line/.style={ultra thick, black}
}
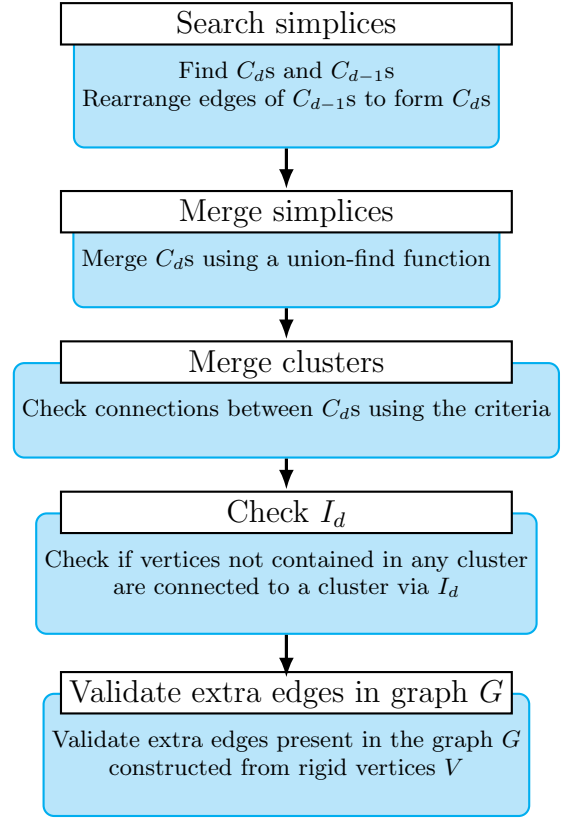
\begin{figure}
    \captionsetup{singlelinecheck=false, justification=raggedright}
    \begin{tikzpicture}[
        node distance=1cm and 1.2cm,
        thick,
        nodes={align=center},
        >={Latex[scale=.9]},
        ortho/install shortcuts]
        \node[basic box=cyan, header=Search simplices, anchor=north] at (0,0) (F1) {Find $C_d$s and $C_{d-1}$s\\
        Rearrange edges of $C_{d-1}$s to form $C_d$s};
        \node[basic box=cyan, header=Merge simplices, anchor=north] at (0,-2.5) (T1) {Merge $C_d$s using a union-find function};
        \node[basic box=cyan, header=Merge clusters, anchor=north] at (0,-4.5) (T2) {Check connections between $C_d$s using the criteria};
        \node[basic box=cyan, header=Check $I_d$, anchor=north] at (0,-6.5) (I1) {Check if vertices not contained in any cluster\\
        are connected to a cluster via $I_d$};
        \node[basic box=cyan, header=Validate extra edges in graph $G$, anchor=north] at (0,-8.9) (I2) {Validate extra edges present in the graph $G$\\constructed from rigid vertices $V$};
        \path[very thick, black] (F1) edge[->] (0,-2.2);
        \path[very thick, black] (T1) edge[->] (0,-4.2);
        \path[very thick, black] (T2) edge[->] (0,-6.2);
        \path[very thick, black] (I1) edge[->] (0,-8.7);
    \end{tikzpicture}
    \caption{{\it Computational processes for identifying rigid clusters.} Note that the second and third steps can be combined into one.}
    \label{fig:alg-flow}
\end{figure}
\begin{figure}
		\captionsetup{singlelinecheck = false, justification=raggedright}
		\begin{center}
\begin{tabular}{cc}
     \includegraphics[width=0.2\textwidth]{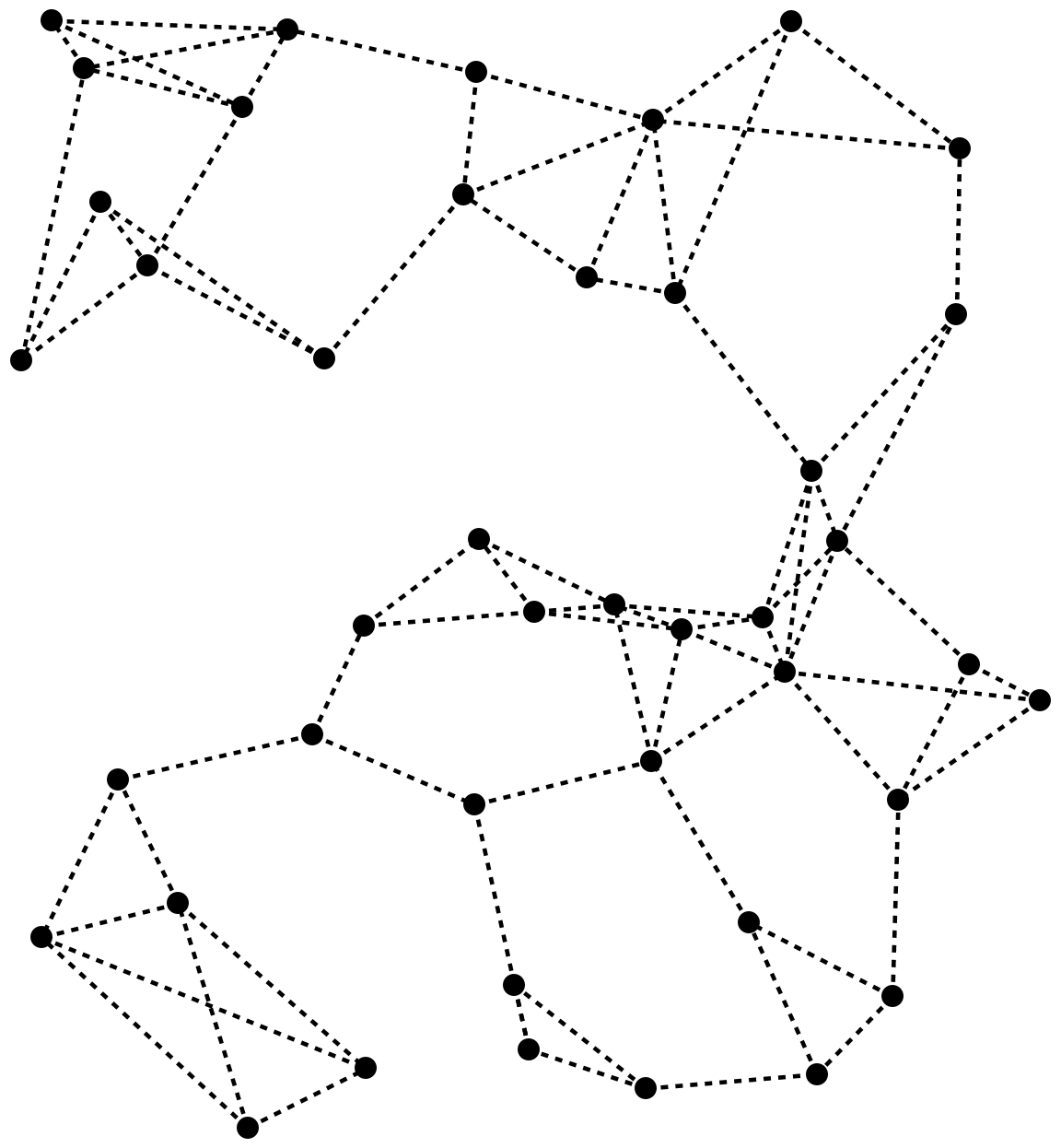}&
     \includegraphics[width=0.2\textwidth]{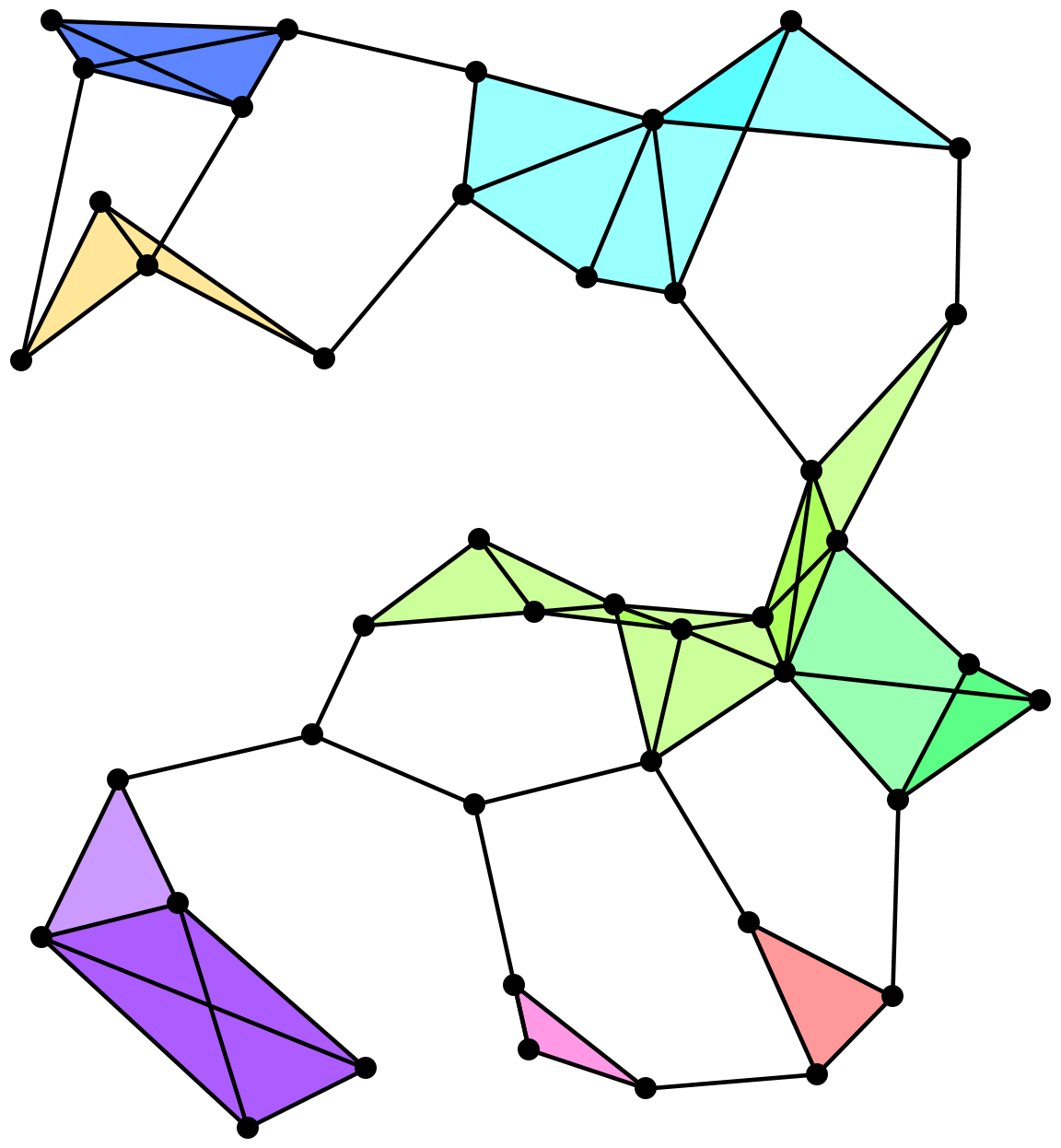}\\
     (a) & (b) \\ 
     \includegraphics[width=0.2\textwidth]{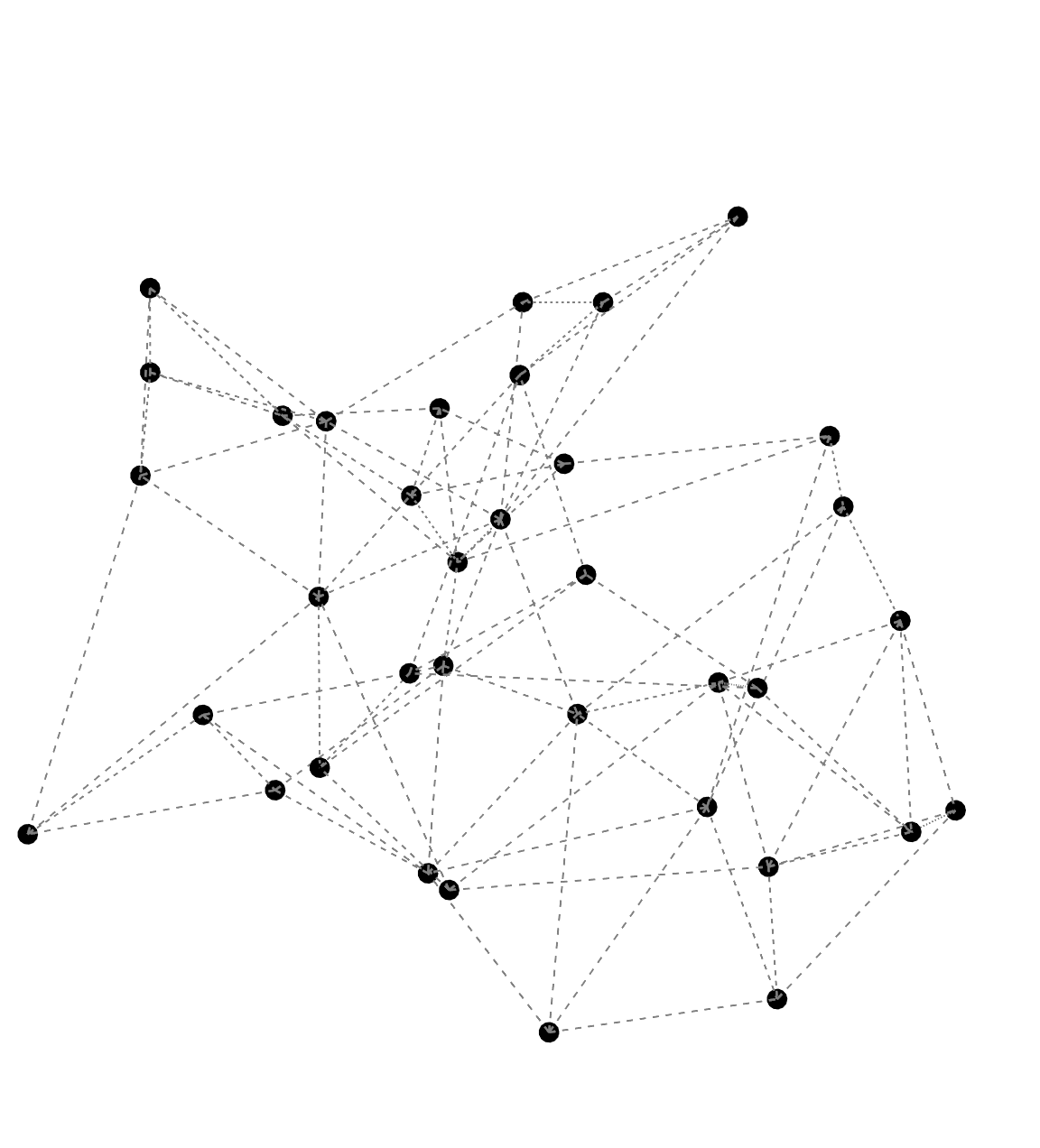}&
     \includegraphics[width=0.2\textwidth]{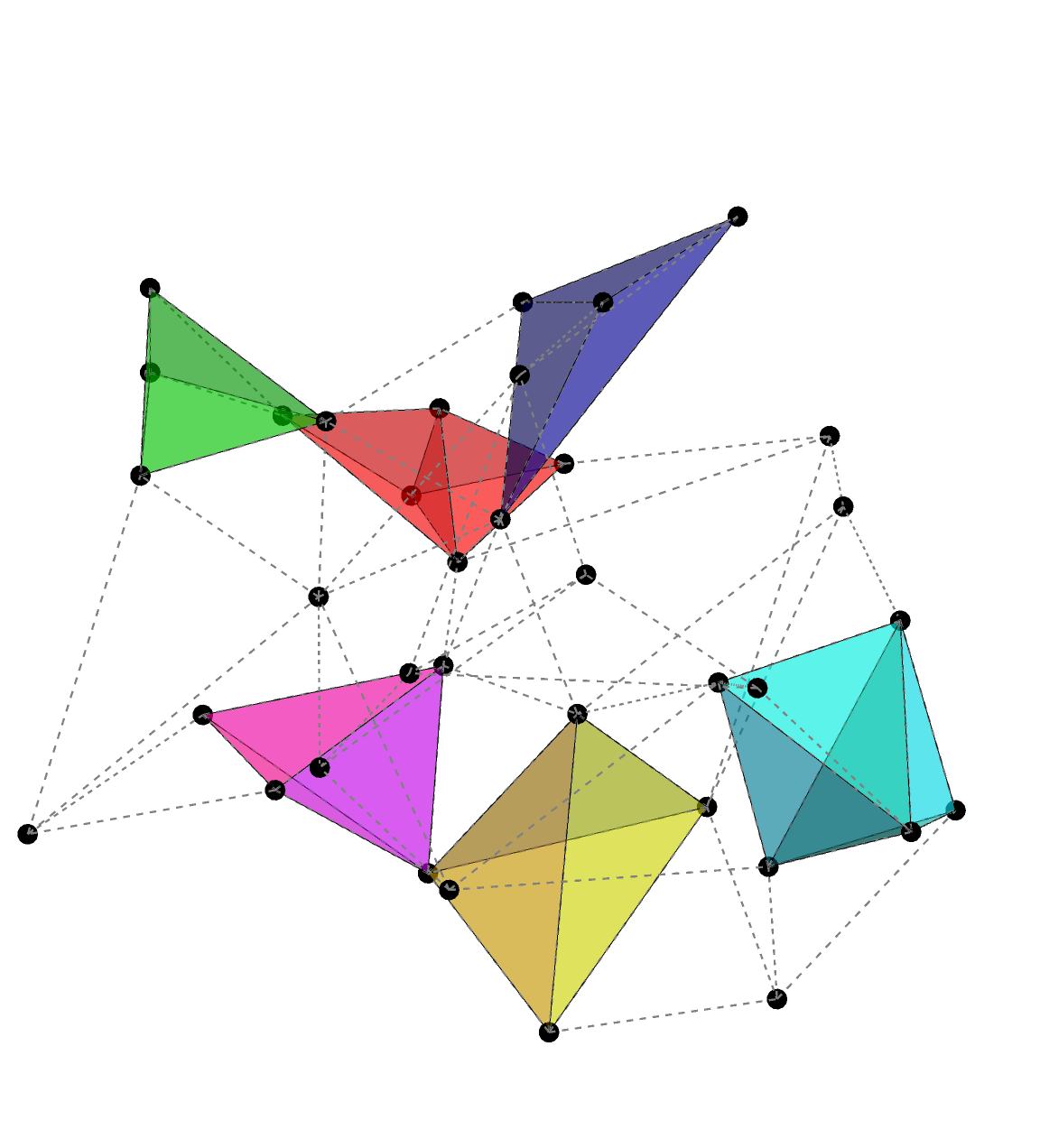}\\
     (c) & (d) \\ 
\end{tabular}
\end{center}\caption{{\it Computational examples in 2D and 3D} (a) 42 randomly generated points in 2D, (b) rigid clusters in 2D after the merging algorithm, represented in different colors, (c) 36 randomly generated points in 3D, (d) rigid clusters in 3D after the merging algorithm, represented in different colors.}
\label{fig:alg-sample}
\end{figure}

\section{Discussion}
We have introduced methods to assess whether a graph is generically rigid in $\mathbb{R}^d$, showing that our counting rules are directly connected to Tay's theorem, a renowned criterion for checking the rigidity of body-bar graphs. Additionally, we explained the concept of pinned rigid graphs using the gluing lemma (Lemma \ref{gluelemma3}), demonstrating that any rigid graph can be interpreted as a combination of pinned points and pinned rigid components. We first illustrated this in 2D and 3D and then extended the concept to $d$-dimensional space with a straightforward proof.

Using these rules, we have outlined a process to determine rigidity by examining the connections within the graph. We then introduced methods for constructing rigid graphs through both Henneberg-type and non-Henneberg-type operations. In the final section, we presented an algorithm that checks graph rigidity based solely on connectivity, leveraging the previous counting rules.

In two-dimensional space, the pebble-game method \cite{JACOBS1997} is a well-established and efficient tool for checking rigidity. A generalized version of this method for $(k,l)$-sparse graphs was developed by Lee, {\it et al.} \cite{LEE2008}, and Lee, {\it et al.} \cite{Lee2005} extended it for planar bar-and-joint graphs and $d$-dimensional body-bar rigidity. The pebble-game method aligns with the 2D counting rules by assigning pebbles to nodes and counting the leftover pebbles, with a complexity of $\mathcal{O}(n^2)$ \cite{LEE2008}. However, in three-dimensional space, the counting rule is not always applicable, as demonstrated by the double-banana graph example.

Several studies have built upon the pebble-game method to three dimensions. For example, Chubynsky {\it et al.} \cite{Chubynsky2007} integrated the (3,6) pebble game with hinge checks and stress analysis. Bruun {\it et al.} \cite{BRUUN2022} developed a 3D algorithm based on the Henneberg sequence from a mechanical perspective, while Bereg \cite{Bereg2005} proposed an algorithm for $d$-dimensional rigidity using the $d$-dimensional rigidity matroid. However, the matrix for large graphs, such as a $10000 \times 10000$ matrix, can consume gigabytes of space in .csv format when coded in Python, presenting challenges for systems with limited computing power. The matrix's sparsity further reduces its efficiency as a storage method.

To address these challenges, we propose a new algorithm based on the simplicial structure of the Henneberg sequence in $d$-dimensional space. Our approach minimizes the use of matrices, such as full-size adjacency and rigidity matrices, while maintaining computational efficiency with a complexity lower than $\mathcal{O}(n^d)$. Specifically, we estimate the following complexities based on Fig. \ref{fig:alg-flow}:
\begin{itemize}
    \item $\mathcal{O}(nd_{\text{max}})$ for searching simplices,
    \item Approximately $\mathcal{O}(2N_{s})$ for merging $N_{s}$ simplices and $N_{c}$ clusters ($N_{s} \geq N_{c}$),
    \item $\mathcal{O}(n_{\text{leftover}})$ for checking $n_{\text{leftover}}$ vertices that do not belong to rigid clusters,
    \item $\mathcal{O}(N_{c})$ for examining extra edges associated with rigid vertices of $N_{c}$ clusters but omitted from the edge list.
\end{itemize}
We anticipate that the overall computation time will be comparable to $\mathcal{O}(n^2)$ across dimensions, excluding the time required for set operations such as $V(G_1) \cap V(G_2)$.

While the rigidity matrix is a robust tool for evaluating rigidity in $d$-dimensional graphs, its feasibility for large graphs is constrained by computing resources due to its size and sparsity. For graphs with a small number of rigid clusters, repeatedly applying the rigidity matrix with different vertex and edge subsets can be inefficient. To address these challenges, our approach reduces reliance on the rigidity matrix by using a connectivity list with a union-find function to enhance computational speed. Additionally, we acknowledge that there are various algorithms for finding simplices beyond those we have discussed. Moreover, as we mentioned earlier, our method can be optimized and enhanced by using a relatively small-sized rigidity matrix inside the algorithm. Therefore, the algorithm can be further improved based on the specific structure of the graph. This approach is adaptable depending on available resources, programming languages, or platforms. In sum, this work lays the foundation for algorithms in multi-dimensional rigidity analysis, facilitating the efficient evaluation of large graphs for many practical applications.

\newpage
\appendix
\section{Important relations}
\setcounter{figure}{0}

\begin{definition}
    The direct sum $\oplus$ of two non-overlapping sets $A$ and $B$ (i.e., $A\cap B = \emptyset$) is defined as $A\oplus B = A \cup B$.
\end{definition}

\begin{definition}[\protect{Chapter B-3}{\cite{RotAlg}}]\label{Matdirsum}
	If $A$ is an $r \times r$ matrix and $B$ is an $s \times s$ matrix, then their $\mathbf{direct}$ $\mathbf{sum}$ $A \oplus B$ is the $(r+s) \times (r+s)$ matrix
	\begin{align}
		A \oplus B = \begin{bmatrix} A & 0 \\ 0 & B \end{bmatrix}.
	\end{align}
\end{definition}

Note that we extend this definition for non-square matrices, where $A$ is an $m \times n$ matrix, $B$ is an $r \times s$ matrix, and $A \oplus B$ becomes an $(m+r) \times (n+s)$ matrix.

\begin{lemma}\label{matrank}
	Let $A$ be an $r \times r$ matrix, $B$ an $s \times s$ matrix, $C$ an $r \times s$ matrix, and $D$ an $s \times r$ matrix. Then:
	\begin{itemize}
		\item[(i)] 	$\rank(A \oplus B) = \rank(A) + \rank(B)$.
		\item[(ii)] $\rank \begin{bmatrix} A & 0 \\ C & B \end{bmatrix} \geq \rank(A) + \rank(B)$.
		\item[(iii)] $\rank \begin{bmatrix} A & D \\ 0 & B \end{bmatrix} \geq \rank(A) + \rank(B)$.
	\end{itemize}
\end{lemma}

\begin{proof}
	\begin{enumerate}
		\item[(i)] Suppose $\rank(A) = a$ and $\rank(B) = b$. Let $P, Q$ be $r \times r$ matrices and $P', Q'$ be $s \times s$ matrices. Define $\widetilde{P} = P \oplus P'$ and $\widetilde{Q} = Q \oplus Q'$. From matrix properties, there exist factorizations $A = PE^{r,r}_aQ$ and $B = P'E^{s,s}_bQ'$. Then:
		\begin{align*}
			A \oplus B &= \begin{bmatrix} PE^{r,r}_aQ & 0 \\ 0 & P'E^{s,s}_bQ' \end{bmatrix} \\
			&= \begin{bmatrix} P & 0 \\ 0 & P' \end{bmatrix} \begin{bmatrix} E^{r,r}_a & 0 \\ 0 & E^{s,s}_b \end{bmatrix} \begin{bmatrix} Q & 0 \\ 0 & Q' \end{bmatrix} \\
			&= \tilde{P}(E^{r,r}_a \oplus E^{s,s}_b)\tilde{Q}.
		\end{align*}
		The matrix $E^{r,r}_a \oplus E^{s,s}_b$ contains the standard basis vectors $\epsilon_1, \dots, \epsilon_a$ from $E^{r,r}_a$ and $\epsilon_{r+1}, \dots, \epsilon_{r+b}$ from $E^{s,s}_b$. Thus, $\rank(A \oplus B) = \rank(A) + \rank(B)$.
		\item[(ii)] Take $A = 0$, $B = 0$, and $C \neq 0$. Since the rank of any matrix is at least 0, we obtain the desired result.
		\item[(iii)] Similar to (ii), take $A = 0$, $B = 0$, and $D \neq 0$.
	\end{enumerate}	
\end{proof}

\begin{lemma}\label{matrank2}
	Let $A$ be an $m \times n$ matrix, $B$ an $r \times s$ matrix, $C$ an $m \times s$ matrix, and $D$ an $r \times n$ matrix. Then:
	\begin{itemize}
		\item[(i)] 	$\rank(A \oplus B) = \rank(A) + \rank(B)$.
		\item[(ii)] $\rank \begin{bmatrix} A & 0 \\ C & B \end{bmatrix} \geq \rank(A) + \rank(B)$.
		\item[(iii)] $\rank \begin{bmatrix} A & D \\ 0 & B \end{bmatrix} \geq \rank(A) + \rank(B)$.
	\end{itemize}
\end{lemma}

\begin{proof}
	The proof follows similarly to Lemma \ref{matrank}.
\end{proof}

\begin{theorem}[\protect{Theorem 2.2}{\cite{Sch10}}]\label{diso}
	For a $d$-dimensional realization $(G,p)$ of a graph $G$ with $|V(G)| \geq d$, the following are equivalent:
	\begin{itemize}
		\item[(i)] $(G,p)$ is isostatic.
		\item[(ii)] $(G,p)$ is infinitesimally rigid, and $|E(G)| = d|V(G)| - \binom{d+1}{2}$.
		\item[(iii)] $(G,p)$ is independent, and $|E(G)| = d|V(G)| - \binom{d+1}{2}$.
	\end{itemize}
\end{theorem}

\begin{lemma}[\protect{Lemma 11.1.9/Lemma 2.1}{\cite{Whiteley95,Jackson05}}]\label{glulemma}
	Suppose $G = G_1 \cup G_2$.
	\begin{itemize}
		\item[(i)] If $|V(G_1) \cap V(G_2)| \geq d$ and both $G_1$ and $G_2$ are rigid in $\mathbb{R}^d$, then $G$ is rigid in $\mathbb{R}^d$.
		\item[(ii)] If $E_1$ and $E_2$ are generically $d$-independent and $E_1 \cap E_2$ is generically $d$-rigid, then $E_1 \cup E_2$ is generically $d$-independent.
	\end{itemize}
\end{lemma}

\begin{lemma}[\protect{Lemma 3.1}{\cite{Jordan2010}}]
    Let $G = (V, E)$ be a graph, and let $P \subseteq V$ with $|P| \geq d$. Then $(G, p)$ is a generic realization of $G$ in $\mathbb{R}^d$ if and only if $G + K(P)$ is rigid in $\mathbb{R}^d$.
\end{lemma}

This lemma shows that both the pin part and the pinned graph part must be rigid. To generalize this concept, we will now rewrite Lemma \ref{glulemma} for a pinned graph.
\begin{lemma}\label{glulemma2}
Suppose $G = G_1 \cup G_2$.
    \begin{itemize}
        \item[(i)] If $G_1(V_1,E_1)$ is rigid and $G_2(V_2,E_2) = G_2(I_2,P_2,E_2)$ is pinned rigid with $|P_2| \geq d$, $V(G_1) \cap V(G_2) = P_2$, and $E_1 \cap E_2 = \emptyset$ in $\mathbb{R}^d$, then $G$ is rigid in $\mathbb{R}^d$.
        \item[(ii)] If $E_1$ is generically $d$-independent and $E_2$ corresponds to edges of a pinned rigid graph $G_2$, which satisfies being generically $d$-independent, and $E_1 \cap E_2 = \emptyset$, then the set $E_1 \cup E_2$ is generically $d$-independent.
        \item[(iii)] If $G_1(V_1,E_1)$ is rigid in $\mathbb{R}^{d-1}$, $G_2(V_2,E_2) = G_2(I_2,P_2,E_2)$ is pinned rigid in $\mathbb{R}^d$, and $V(G_1) = P_2$, then at least $|E_2| = d \cdot |I_2| + |P_2| - d$ is required for $G$ to be infinitesimally rigid in $\mathbb{R}^d$.
        \item[(iv)] Any pinned rigid graph $G_1(V_1,E_1) = G_1(I_1,P_1,E_1)$ can be embedded in a rigid graph $G$ by adding a rigid graph $G_2(V_2,E_2)$ of the same dimension, such that $V(G_1) \cap V(G_2) = P_2$ for $|P_2| \geq d$.
    \end{itemize}
\end{lemma}

\begin{proof}
    \begin{itemize}
        \item[(i)] Since edges connecting unpinned points are not overlapping, we can view this as a direct sum of two rigidity matrices. From Lemma \ref{matrank2}(ii), we have $\rank \mathcal{R}(G,p) \geq \rank \mathcal{R}(G_1,p) + \rank \mathcal{R}(\widetilde{G_2},p) = d \cdot |V(G_1)| - \binom{d+1}{2} + d \cdot |V(G_2)| = d \cdot |V(G)| - \binom{d+1}{2}$. Since $\rank \mathcal{R}(G,p) \leq S(n,d) = d \cdot |V(G)| - \binom{d+1}{2}$, by the Squeeze Theorem, we obtain $\rank \mathcal{R}(G,p) = d \cdot |V(G)| - \binom{d+1}{2}$.
        \item[(ii)] Similar to part (i).
        \item[(iii)] From Theorem \ref{diso}, we require $|E(G)| = d \cdot |V(G)| - \binom{d+1}{2}$. We know that $\rank \mathcal{R}(G_1,p) = (d-1) \cdot |P_2| - \binom{d}{2}$ and $\rank \mathcal{R}(\widetilde{G_2},p) = d \cdot |I_2|$. Since, from Lemma \ref{matrank2}, we have $\rank \mathcal{R}(G,p) \geq \rank \mathcal{R}(G_1,p) + \rank \mathcal{R}(\widetilde{G_2},p)$, and $(d-1) \cdot |P_2| + d \cdot |I_2| - \binom{d}{2} = d \cdot |V(G)| - \binom{d+1}{2} - |P_2| + d$, we get $d \cdot |V(G)| - \binom{d+1}{2} - |P_2| + d \leq \rank \mathcal{R}(G,p) \leq d \cdot |V(G)| - \binom{d+1}{2}$. Therefore, to make $G$ rigid, $|E_2| = d \cdot |I_2|$ is not sufficient, and we require at least $|P_2| - d$ additional edges.
        \item[(iv)] Same as part (i).
    \end{itemize}
\end{proof}

The following theorem is a gluing lemma (Lemma \ref{glulemma}) for a globally rigid graph.

\begin{theorem}[\protect{Theorem 63.2.4}{\cite{Dgeom}}]\label{gluelemma3}
    If $G_1 = (V_1,E_1)$ and $G_2 = (V_2,E_2)$ are globally rigid graphs in $\mathbb{R}^d$ sharing at least $d+1$ vertices, then $G = (V_1 \cup V_2, E_1 \cup E_2 - G_1[V_1 \cap V_2])$ is globally rigid in $\mathbb{R}^d$. If $G_1 = (V_1;E_1)$ and $G_2 = (V_2,E_2)$ are globally rigid graphs in $\mathbb{R}^d$ sharing exactly $d+1$ vertices and some edge $e$, then $G = (V_1 \cup V_2, E_1 \cup E_2 - e)$ is globally rigid in $\mathbb{R}^d$.
\end{theorem}
\section*{Funding}
\section*{Acknowledgements}
JMS would like to acknowledge financial support from the National Science Foundation under grant DMR-2204312.
\newpage
 
\bibliography{main}

\end{document}